\documentclass[twocolumn]{aastex63}
\usepackage{hyperref}
\usepackage{scrextend}
\deffootnote[0.45em]{0.2em}{0.2em}{\textsuperscript{\thefootnotemark}\,}
\usepackage{amsmath,url}
\usepackage{algorithm}
\usepackage{amsfonts}
\usepackage{mathrsfs}
\usepackage{bm}
\usepackage{rotating}
\usepackage{color}
\usepackage{graphicx}
\usepackage{subfigure}
\usepackage{lineno}
\usepackage[toc,page]{appendix}

\title{A multi-epoch X-ray study of the nearby Seyfert 2 galaxy NGC 7479: Linking column density variability to the torus geometry}

\newcommand{\swift}{\textit{Swift}}
\newcommand{\chandra}{\textit{Chandra}}
\newcommand{\xmm}{XMM-\textit{Newton}}
\newcommand{\nustar}{\textit{NuSTAR}}
\def\bat{{{\it Swift}-BAT  }}

\newcommand{\mytorus}{\texttt{MYTorus}}
\newcommand{\uxclumpy}{\texttt{UXCLUMPY}}
\newcommand{\borus}{\texttt{borus02}}

\newcommand{\xspec}{\texttt{XSPEC }}

\shorttitle{Multi-epoch study of NGC 7479's X-ray torus properties }
\shortauthors{Pizzetti et al.}

\begin{document}
\title{A multi-epoch X-ray study of the nearby Seyfert 2 galaxy NGC 7479: Linking column density variability to the torus geometry}

\author{A. Pizzetti}
\affiliation{Department of Physics and Astronomy, Clemson University,  Kinard Lab of Physics, Clemson, SC 29634, USA}

\author{N. Torres-Alb\`{a}}
\affiliation{Department of Physics and Astronomy, Clemson University,  Kinard Lab of Physics, Clemson, SC 29634, USA}

\author{S. Marchesi}
\affiliation{Department of Physics and Astronomy, Clemson University,  Kinard Lab of Physics, Clemson, SC 29634, USA}
\affiliation{INAF - Osservatorio di Astrofisica e Scienza dello Spazio di Bologna, Via Piero Gobetti, 93/3, 40129, Bologna, Italy}

\author{M. Ajello}
\affiliation{Department of Physics and Astronomy, Clemson University,  Kinard Lab of Physics, Clemson, SC 29634, USA}

\author{R. Silver}
\affiliation{Department of Physics and Astronomy, Clemson University,  Kinard Lab of Physics, Clemson, SC 29634, USA}

\author{X. Zhao}
\affiliation{Center for Astrophysics | Harvard \& Smithsonian, 60 Garden Street, Cambridge, MA 02138, USA}

\begin{abstract}
\noindent Active galactic nuclei (AGN) are powered by accreting supermassive black holes, surrounded by a torus of obscuring material. Recent studies have shown how the torus structure, formerly thought to be homogeneous, appears to be ‘patchy’: the detection of variability in the line-of-sight hydrogen column density, in fact, matches the description of an obscurer with a complex structure made of clouds with different column density.
In this work, we perform a multi-epoch analysis of the X-ray spectra of the Seyfert 2 galaxy NGC 7479 in order to estimate its torus properties, such as the average column density and the covering factor. The measurement of the line-of-sight hydrogen column density variability of the torus allows us to obtain an upper limit on the cloud distance from the central engine.
In addition, using the X-ray luminosity of the source, we estimate the Eddington ratio to be in a range of $\lambda_{\rm Edd}=0.04-0.05$ over all epochs.

\end{abstract}

\section{Introduction} 
Active Galactic Nuclei (AGN) are accreting supermassive black holes (SMBHs) at the center of galaxies, surrounded by a structure of obscuring material, which historically has been assumed to have a toroidal shape \citep[unification model,][]{antonucci_1993, urry_padovani}. 

Results obtained by using millimeter/submillimeter, infrared, optical and X-ray facilities \citep[see, e.g.,][]{simpson2005, honig2019, Combes2019, zhao_2021} suggest the presence of a doughnut-like structure of molecular gas and dust on the 0.1-100\,pc scale as the origin of the obscuration and reflection of the radiation coming from the inner part of the AGN. The shape and the size of the torus is still a matter of debate, with recent studies suggesting  a  warped disk-like structure as a more appropriate description of the geometry the obscuring material \citep[]{antonucci_1993, combes2014, Combes2019, audibert2019, buchner2021}.  The radiation is reprocessed by the dust and the molecular gas of the torus; the dust absorbs the radiation and thermally re-emits it in infrared. The thick molecular gas, instead, reprocesses the incoming radiation via photoelectric absorption and Compton scattering, generating the Compton hump feature present n the X-ray spectra of the most obscured sources. 


Infrared and X-ray studies of the torus \citep[see, e.g.,][]{Risaliti2002,risaliti_2005, bianchi_2005, risaliti_2011, Sanfrutos2013, Markowitz_2014, Balokovic2018, Buchner_2019, Laha2020} have ruled out the initial idea of a homogeneous torus, leaving the clumpy scenario, in where the torus is made of multiple clouds of different densities, to be a more accurate description of the the spatial distribution of the obscuring material.
If this is indeed the case, in X-ray we should expect to see line-of-sight (l.o.s.) hydrogen column density (N$_{\rm H,los}$) variability on time scales that vary from days \citep[]{risaliti_2005}, to months \citep[]{Risaliti2002}, to years \citep[]{Elvis_2004}, depending on the size of the clouds and their distance from the X-ray emitter \citep[]{risaliti_2005, nenkova2008}. 

By studying the obscuration variability of a large sample of Seyfert 1 and Compton-thin Seyfert 2 galaxies ($N_{H,los}<1.5\times10^{24}$\,cm$^{-2}$)\footnote{Based on their optical spectra, Seyfert galaxies have been classified Seyfert 1, when both narrow and broad line region are visible and Seyfert 2, when only the narrow lines are present, due to the obscuration of the torus \citep[][]{urry_padovani,antonucci_1993}.}, \citet{Markowitz_2014} and \citet{Laha2020} have deeply investigated the clumpy-torus scenario.  A major result from these studies is the construction of a solid X-ray based statistical support for it; the two works, in fact, show how eclipsing events are common among Sy2 galaxies and how a scenario in where clouds of different densities and radial velocities moving inside a more diffuse medium is preferred over the classic homogeneous donut-shaped torus one.  

For this work, we focused on the analysis of a single source, NGC 7479, having as main goal the characterization of the properties and the geometry of its torus. The study was done by analyzing the hydrogen column density variability via a simultaneous fit of multi-epoch X-ray spectra. Previous works done by \citet{Marchesi_2019} and \citet{zhao_2021} showed this galaxy to be a heavily obscured AGN \citep[{line-of-sight hydrogen column density N$_{\rm H,los}=24.76^{+0.08}_{-0.07}$\,cm$^{-2}$,}][]{zhao_2021} and a potential N$_{\rm H,los}$ variable AGN. As shown in \citet{Balokovi__2019} and \citet{saha_2022}, the multi-epoch X-ray spectral analysis is a powerful tool to constrain torus parameters such as the covering factor, the torus inclination angle, the photon index and the average hydrogen column density, while, at the same time, analyzing the N$_{\rm H,los}$ variability.

The paper is structured as follows: in Sections \ref{sec:Observ} and \ref{sec:analysis}, the data reduction and analysis is performed, as well as the description of the models used for the spectral fitting. In Section \ref{sec:results} we discuss the results obtained in this work and compare them to previous analysis of NGC 7479; in particular, we focus on the connection between the line-of-sight hydrogen column density variability with respect to the torus geometry. Finally, we report our conclusions in Section \ref{sec:conclusion}.
The cosmological parameters assumed in this work are compatible with a flat $\Lambda$CDM cosmology with H$_0$=69.6\,km\,s$^{-1}$\,Mpc$^{-1}$, $\Omega_m$=0.29 and $\Omega_\Lambda$=0.71 \citep{bennett2014}. All reported uncertainties are at 90\% confidence level unless otherwise stated.



%
%
\section{Observations and Data Analysis}\label{sec:Observ}
NGC 7479  \citep[$z\sim 0.007941$, $d\sim 36$\,Mpc,][]{haynes1998} is a Seyfert 2 galaxy \citep[]{Lumsden_2001} detected in the 150-month BAT catalog (Imam et al. in prep.\footnote{The online version of the catalog can be found at \url{https://science.clemson.edu/ctagn/bat-150-month-catalog/}}), a catalog of $\sim$1000 AGN detected by \bat in the 15-150\,keV range. 
The source was selected from the work of \citet{zhao_2021} because of the number and cadence of its existing observations, and because of the tentative detection of line-of-sight N$_{\rm H}$ variability in the mentioned work. The analysis of one \nustar\ (2016-05-12) and one \xmm\ (2018-05-30) observations reported in \citet{zhao_2021}, show a $\Delta log$ N$_{\rm H,los}=0.20_{-0.08}^{+0.10}$ between the line-of-sight column densities of the two observations, in a time-span of 2.05 yr.
Further public archival data are available in the 0.5-10\,keV and 3-70\,keV range for \xmm/\chandra\  and  \nustar, respectively. In total, NGC 7479 has two observations with \chandra, one with \nustar, and three with \xmm. Table \ref{table:sources} reports the main details of the observations analyzed in this work.

\begin{figure*}[ht]
    \centering
    \includegraphics[width=1.0\textwidth]{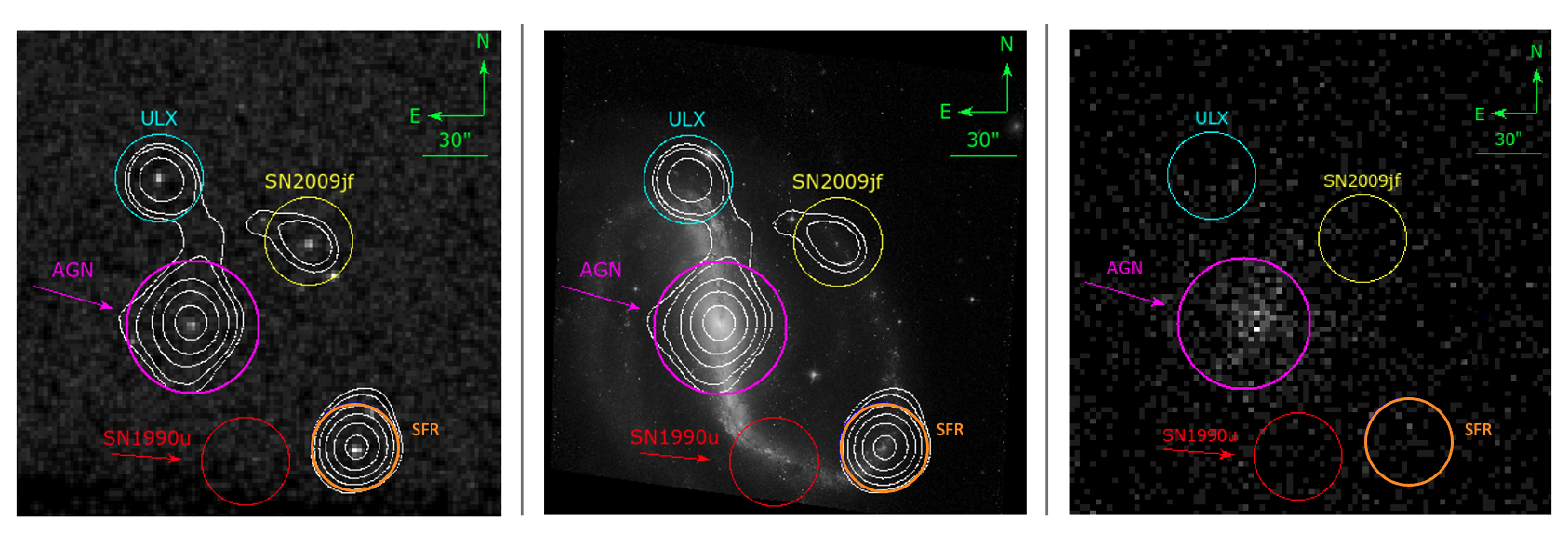}
    \caption{\xmm\ EPIC-PN contours (white) overplotted on \chandra\ - b (left) and \textit{Hubble Space Telescope} (center) observation of NGC 7479. The right panel shows the \nustar\--1 observation. In the figure, the other X-ray sources in the galaxy are also highlighted: the ULX (cyan), the SFR (orange), the supernova SN2009jf (yellow), the supernova remnant of SN1990u (red) and the central engine (magenta). The regions showed in the Figure are for display only and do not correspond to the ones used for the spectral extraction.}
    \label{fig:xmm_chandra}
\end{figure*}

NGC 7479 contains multiple X-ray sources, as can be seen in the $\sim$3$^\prime\times$3$^\prime$ region in Figure \ref{fig:xmm_chandra}. The galaxy has been the host of two supernova explosions  \citep[SN1990u][and SN2009jf, \citealt{Kasliwal_2009}]{Pennypacker_1990}, the latest of which was detected by \chandra. The high resolution images obtained with \chandra\ allowed us to highlight the presence of an Ultra-Luminous X-ray source \citep[ULX, ][]{voss_2011}, a star forming region \citep{zhou_2011} and the supernova SN1990u \citep{Kasliwal_2009}. Figure \ref{fig:xmm_chandra} - left, also shows \xmm\ intensity contours overlaid on the \chandra\--b image, which indicates that the separation between the sources is large enough to isolate the AGN from the other elements, not only in the \chandra\ image, but also in the \xmm\ one. The \nustar\ images show one single source, as can be seen in Figure \ref{fig:xmm_chandra} - right. 
To minimize any possible residual contamination, however, we limited the \nustar\ extraction radius to $30"$, same as used for \xmm.



\begin{table*}
\centering
\vspace{.1cm}

  \begin{tabular}{lcccc}
 
       \hline
       \hline
    \textbf{Instrument}  & \textbf{ObsID}     &    \textbf{Exposure Time  [ks]} &   \textbf{Start Date}  \\ 
    \hline
     \xmm\ - a       &   0025541001   &   5.6            &   2001-06-19 \\
     
      \xmm\  - b       &   0301651201   &   16.0            &   2006-07-13\\
    
    \textit{Chandra} - a   &   10120   &     25.1               &   2009-08-11 \\
    
    \textit{Chandra} - b    &   11230   &     10.1               &   2009-10-24 \\
    
    \nustar\- 1     &   60201037002   &  18.5           &   2016-05-12 \\
    
    \xmm\  - c        &   0824450601   &   64.4           &   2018-05-30 \\
    
    \nustar\- 2    &   60061316002   &  23.6           &   2020-11-06 \\
    
       \hline
     
\end{tabular}
 \caption{Source observation details of NGC 7479. For \xmm, the reported exposure times correspond to EPIC-PN (after removing background flares). The \xmm\-- b observation was removed due to flare contamination.}
  \label{table:sources}
\vspace{.2cm}
\end{table*}

\subsection{\chandra\  data reduction}
For this work, we use two \chandra\ observations taken on 2009 August 11 and 2009 October 24, with an exposure time of $\sim$25\,ks and $\sim$10\,ks respectively. 


The data were reduced with the CIAO \citep[v4.14;][]{fruscione2006} software and the \chandra\ Calibration Data Base \texttt{caldb} 4.9.7 adopting standard procedures. The source and the background spectra were extracted using the CIAO \texttt{specextract} tool. For the source, we selected a circle of $5"$, while the background was extracted from an annulus of internal radius $r_{in}=6"$ and external radius $r_{out}=15"$, after a visual inspection to avoid the presence of any other source in the field. Due to the low exposure time, we grouped each spectrum with a minimum of 5 counts per bin with the \texttt{grppha} tool. Therefore, the spectra are fitted with the \texttt{cstat} statistics \citep[]{cash1979}.

\subsection{\xmm\ data reduction}
NGC 7479 has three archival \xmm\ observations taken on 2001 June 19, 2006 July 13 and 2018 May 5, but only two of them are included in this analysis (see Table \ref{table:sources}), as the second \xmm\ observation (\xmm\--b in Table \ref{table:sources}) is heavily contaminated by flares. The two remaining observations, of 5.6\,ks and 64.6\,ks respectively, were reduced using the Science Analysis System \citep[SAS;][]{jansen2001} version 19.0.0. In order to remove times when the particle background was high, we inspected the lightcurve at energies E$>$10\,keV. Then, we chose a value of 0.35 counts/s for both MOS1 and MOS2, and 0.4\,cts/s for EPIC-pn to remove bright background flares visible in both observations. 

For the extraction region, we selected a region of 30" for the source and of 40" for the background for all the detectors. Considering the multiple X-ray sources (see Figure \ref{fig:xmm_chandra}), we visually inspected the images to avoid any contamination. 
Finally, we binned the spectra to have at least 15 counts per bin, in order to use the $\chi^2$ statistics.

\subsection{\nustar\ data reduction}
NGC 7479 has been observed by \nustar\ twice, on 2016 March 12 and on 2020 November 6, with an exposure time of $\sim$ 18\,ks  and $\sim$ 23 ks respectively. The \nustar\ data were retrieved from both focal plane modules, FPMA and FPMB. The event data file were calibrated and cleaned using the \nustar\ \texttt{nupipeline} script version 0.4.8. and the Calibration Database (CALDB) v.20210427 as response file. Then, we used the \texttt{nuproducts} script to generate the  ARF, RMF and light-curve files. For both observations, the source spectrum was extracted from a 30" circular region centered on the source's optical position. 
We then extracted the background spectrum from each module, choosing a circular region of $40"$ after a visual inspection to avoid contamination from any other sources. Lastly, the \nustar\ spectra were grouped with the \texttt{grppha} task with a minimum of 15 counts per bin and the $\chi^2$ statistics was used.

%
%

\section{X-ray spectral analysis}\label{sec:analysis}

In the following section we describe the different torus models used in the spectral analysis of NGC 7479. The results of this analysis will be shown in Section \ref{sec:results}.

To all models, we add a thermal emission component \citep[\texttt{apec},][]{smith2001} and two emission lines (\texttt{zgauss}), in order to account for the soft X-ray emission from the central region of the host galaxy. A multiplicative constant is added before the \texttt{apec} component ($C_{\rm apec}$) in order to have a better modulation of the emission between different observations.  The absorption along the line of sight due to our Galaxy is taken into account by including the \texttt{phabs} component ($4.84\times 10^{20}$\,cm$^{-2}$) into the models \citep[]{Kalberla2005}. Finally, we model the fraction of the intrinsic AGN power law that is scattered without being reprocessed by the obscuring material, by multiplying the intrinsic power law by the fraction of the scattered emission, $F_s$.

We apply a multiplicative constant, $C_{\rm AGN}$, to the intrinsic power-law model, to disentangle intrinsic flux variability from column density variability between the various observations.


We analyzed the torus properties of NGC 7479 by using three models based on Monte-Carlo simulations that can self-consistently describe the primary AGN emission with the Compton-thick gas (N$_{\rm H,los}>1.5\times 10^{24}$\,cm$^{-2}$) in the surrounding torus. The physical models used in this work are \borus\ \citep[]{Balokovic2018}, \mytorus\ \citep[]{Murphy2009, Yaqoob2012, yaqoob2015},  and \uxclumpy\ \citep[]{Buchner_2019}.

In \texttt{XSPEC} \citep{Arnaud1996} this configuration is as follows:

\begin{equation}
\begin{split}
Model =  phabs &* \{C_{\rm apec}*(apec+zgauss+zgauss) \\
&   +C_{\rm AGN}*(TorusModel +F_s*pwl)\}
\end{split}
\end{equation}


For consistency, we fit all the models in the 0.6-50\,keV range. 
In the simultaneous fitting of all the observations, only $C_{\rm AGN}$ and N$_{\rm H,los}$ are left free to vary between the observations, as we have to take into account possible flux variations and we expect variability in the line-of-sight hydrogen column density. The models' capabilities of disentangling between the line-of-sight column density variability and the intrinsic flux variability is shown by the confidence contour plots (Fig. \ref{fig:cont_plots}) in Appendix \ref{appendix:cont_plots}. Other parameters such as the covering factor or the average column density of the torus are instead linked between the observations, as they refer to global torus properties that we do not expect to change over the time-scales sampled in this work.
As we found the high-energy cutoff to be an unconstrained parameter in our analysis, and in order to be more consistent with recent estimations of its median value \citep[{$\sim 300$\,keV,}][]{balokovic2020}, we froze the high-energy cutoff for all the three models to 300\,keV.



\subsection{borus02}\label{subsec:borus02}

The \borus\ model assumes a uniform-density sphere with two conical cutouts filled with a cold, neutral and static medium. The elemental abundance is assumed to be solar apart from the abundance of iron, which is a free parameter. The line of sight inclination angle is a variable parameter of the model and it can vary between $\theta \in [18.2-87.1^{^{\circ}}]$. In this model, the fraction of the sky covered by the torus as seen from the central engine, also known as the covering factor $C_F$, is left free to vary within the range $C_F \in [0.1-1]$; a small value indicates a disk-like torus or a non-uniform material distribution, while a large value implies a more spherically symmetric torus. From this we can compute the torus half opening angle $\theta_{tor}=arccos(C_F)$.
 In addition, \borus\ allows the independent calculation of the hydrogen line-of-sight column density (N$_{\rm H,los}$) and the average column density of the torus (N$_{\rm H,av}$). 


Since \borus\  models only the reflection component of the AGN emission, which accounts for the continuum and the lines, the absorbed and the scattered components need to be added manually.

In \xspec the model has the following configuration:

\begin{eqnarray}
 \label{eq:borus02}
    TorusModel = \text{borus02\_v170323a.fits}+ \nonumber\\ 
     +zphabs*cabs*cutoffpl.
\end{eqnarray}
where \texttt{zpbabs} and \texttt{cabs} are the photoelectric absorption due to the cold medium and the Compton scattering losses along the line of sight, respectively.



\subsection{MYTorus}\label{subsec:myt}
The \mytorus\ model considers a cylindrical, azimuthally symmetric torus with a fixed half-opening angle of $60^{^{\circ}}$, filled with a uniform neutral cold reprocessing material. In this model, the main components of an obscured AGN X-ray spectrum (the line of sight, the reflection and the emission line component) are treated self-consistently by the use of three different tables, as can be seen in Eq.~\eqref{eq:myt}. In particular, the Compton-scattered and the lines components are weighted differently by the addition of multiplicative constants $A_S$ and $A_L$, respectively.

In this work, we use the decoupled configuration of this model \citep{Yaqoob2012, yaqoob2015}, as the coupled configuration \citep{Murphy2009} does not allow to disentangle the line of sight column density from the average one (N$_{\rm H,av}$). 
In the decoupled configuration, the zeroth-order continuum (the continuum photons that escaped the torus without being scattered) is independent from the inclination angle, that is fixed to be $\theta_i=90^{\circ}$. In this way, the zeroth-order continuum is independent of geometry and it becomes purely a line-of-sight quantity. 
In order to take into account the possible patchiness and configurations of the torus and of the consequent Compton-scattering and lines features, these two components are considered both in an edge-on and face-on configuration. In the first case the inclination angle, set to be $\theta_{i,S,L}=90^{^{\circ}}$, mimics the forward scattering and it is weighted by $A_{S,L90}$; this means that we are accounting for a more uniform torus, as the photons are primarily reprocessed by the obscuring material that is lying between the AGN and the observer.
In the second case, $\theta_{i,S,L}=0^{^{\circ}}$ accounts for a backward scattering and $A_{S,L0}$ is the weighting constant. This second scenario is more likely to happen when the torus presents a more patchy structure, in which the photons scattered by the back side of the torus have less chance to interact again with the material before reaching the observer. When $A_{S,L90}$ and $A_{S,L0}$ are left free to vary, we refer at the configuration as "decoupled free" and a ratio between the two constants can give a qualitative idea of which emission is more prominent, thus giving us an indication of the inclination angle of the torus. In addition, a ratio between N$_{H,los}/$N$_{H,av}$ can give an approximate estimation of the clumpiness of the torus.

In \xspec the model is as follows:
\begin{eqnarray}
 \label{eq:myt}
  TorusModel =  \text{mytorus\_Ezero\_v00.fits} * zpowerlw + \nonumber \\ 
  +A_{\rm S,0} * \text{mytorus\_scatteredH300\_v00.fits} + \nonumber \\ 
  +A_{\rm L,0} * \text{mytl\_V000010nEp000H300\_v00.fits} +\nonumber \\ 
  +A_{\rm S,90} * \text{mytorus\_scatteredH300\_v00.fits} + \nonumber \\ 
  +A_{\rm L,90} * \text{mytl\_V000010nEp000H300\_v00.fits}.
\quad
\end{eqnarray}


\subsection{UXCLUMPY}\label{subsec:uxclumpy}

\uxclumpy\ \citep[]{Buchner2019} is a physically motivated model constructed to reproduce and model the column density and the cloud eclipsing events in AGN tori in terms of their angular sizes and frequency. In this model, an additional Compton-thick reflector near the corona is added to model strong reflection features in a more precise way.

One of the main differences between \uxclumpy\ and the models described before consists in the inclusion of the clumpiness and the dispersion of the clouds in the model. The model is constructed to reproduce a cloud distribution with different hydrogen column density based on eclipse event rates \citep[]{Buchner2019}, assuming for the clouds circular Keplerian orbits on random planes for simplicity. The dispersion of such distribution is modulated by \texttt{TORsigma} ($\sigma \in [6-90^{\circ}]$), where a large value stands for a large dispersion of the clouds. In some cases, an inner ring of homogeneous  gas can be added and modulated by \texttt{CTKcover} (C $\in [0-0.6]$); the modeling of the dimension and the extension of the clouds allows to accurately reproduce the reflection hump \citep[see, e.g.,][]{Buchner_2019}.


The line of sight inclination angle $\theta_{inc}$ is left free to vary in the range $[0-90^{\circ}]$. Differently from what is seen in the previous models, \uxclumpy\ does not allow to distinguish between N$_{\rm H,los}$ and N$_{\rm H,av}$, as the only component taken into account in the model is a total line-of-sight column density that can  vary between N$_{\rm H,los}\in [10^{20-26}$\,cm$^{-2}]$.

In \texttt{XSPEC}, the model is as follows:

\begin{eqnarray}
 \label{eq:uxclumpy}
    TorusModel = \text{uxclumpy-cutoff.fits}+ \nonumber\\ +F_s*\text{uxclumpy-cutoff-omni.fits}.
\end{eqnarray}
The first table includes the torus transmitted and reflected component with fluorescent emission lines while the second one, multiplied by the scattering fraction, takes into account the presence of a \textit{warm mirror emitter}, a volume-filling gas between the clumps that is, in part, responsible for the scattering of the intrinsic AGN powerlaw.



\section{Results and Discussion}\label{sec:results}
In this section, we present the results of the spectral analysis of NGC 7479. We analyze \chandra\ and \xmm\ spectra in the 0.6-8\,keV range, while for \nustar\ spectra we select the 3-50\,keV range, since the background contribution becomes dominant at higher energies. As mentioned before, we include a thermal component (\texttt{apec}) to the physically motivated models in order to better constrain the soft emission of the galaxy\footnote{The thermal component considered here does not take into account multiple gas phases, locations or temperatures, that should be taken into account in order to get a better model of the soft X-ray emission of the host galaxy.}; nevertheless our data resolution did not allow us to make further modeling of the gas properties in the galaxy \citep[see, e.g., ][]{torres-alba_2018}. The addition of two Gaussian lines was also required at energies E $\sim$ 0.68\,keV and 1.31\,keV in order to reproduce the soft-band emission: these line energies are compatible with those of F$K\alpha 1$ and Mg$K\beta 1$ emission lines. The best-fit values of the simultaneous multi-epoch analysis of NGC 7479 X-ray spectra are shown in Tables \ref{table:fitting}, \ref{table:nh_values}, while a more detailed version can be seen in Table \ref{table:allbestfit} in the Appendix \ref{sec:allbest}. The single-epochs X-ray spectra of NGC 7479 are show in Appendix \ref{appendix:single_ep}.

\subsection{borus02}\label{subsec:borus_results}
\begin{figure}[ht]
    \centering
    \includegraphics[trim={0cm 1cm 0 0cm},clip,width=0.5\textwidth ]{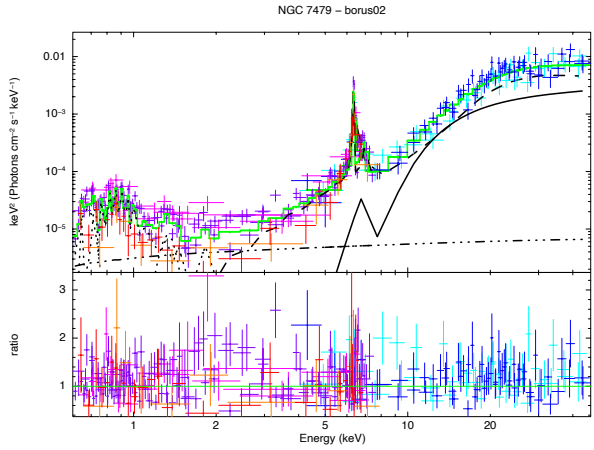}
    \caption{Unfolded \chandra\ (orange and red), \xmm\ (purple and magenta), and \nustar\ (blue and cyan) 0.6-50\,keV combined spectra of NGC 7479 modeled with \borus. The best-fit model is plotted with a solid green line, while the individual model components are plotted in black. Line-of-sight emission: solid line. Reflection component: dashed line. Scattering component: dash-dotted. \texttt{apec}: dotted line.}
    \label{fig:borus_spectrum}
\end{figure}

As already mentioned in Section \ref{subsec:borus02}, in this model the average column density and the covering factor of the torus are free parameters. To account for the line-of-sight column density, following \citep{Balokovic2018}, we use \texttt{zphabs*cabs} to properly measure the Compton scattering. The best-fit values for this models show a $\Gamma=1.79_{-0.06}^{+0.01}$ and a  N$_{\rm H,av}=17.3_{-11.7}^{+30.9}\times10^{24}$\,cm$^{-2}$, both in accordance with what was found by \citet{Marchesi_2019}. The covering factor is $C_F=0.88_{-0.01}^{+0.01}$, corresponding to a half opening angle of the torus $\theta_{\rm tor}=arccos(C_F)=28.35_{-1.19}^{+1.22}$, while the inclination angle, left free to vary, results to be $\theta_i=37.8_{-0.9}^{+0.9}$. The best-fit model and the combined spectra are shown in Figure \ref{fig:borus_spectrum}.

\subsection{MYTorus decoupled}\label{subsec:myt_decf_results}
\begin{figure}[ht]
    \centering
    \includegraphics[trim={0cm 1cm 0 0cm},clip,width=0.5\textwidth ]{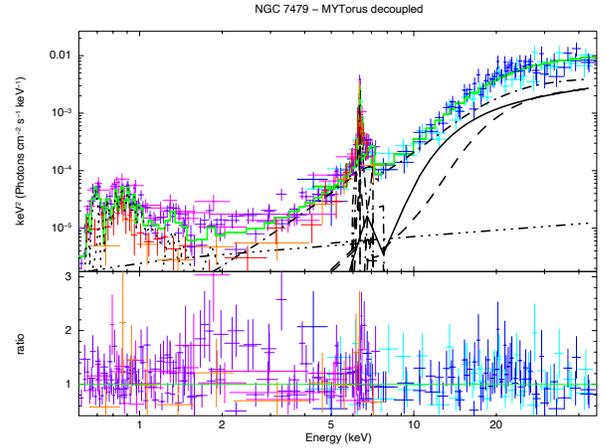}
    \caption{Unfolded \chandra\ (orange and red) \xmm\ (purple and magenta) and \nustar\ (blue and cyan) 0.6-50\,keV combined spectra of NGC 7479 modeled with \mytorus. The best-fit model is plotted with a solid green line, while the individual model components are plotted in black. Line-of-sight emission: solid line. Reflection component:  $90^{\circ}$ reflection is plotted with dashed line, while the $0^{\circ}$ reflection with a dash-dot-dash line; Scattering component: dash-dotted. \texttt{apec}: dotted line.}
    \label{fig:myt_spectrum}
\end{figure}

As mentioned in Section \ref{subsec:myt}, we use the \mytorus\ model in the \textit{decoupled configuration}, meaning that the constant $A_{S,90}$ and $A_{S,0}$ are left free to vary. The best-fit model and the spectra are presented in Figure \ref{fig:myt_spectrum}. The best-fit photon index $\Gamma=1.69_{-0.10}^{+0.05}$ and the N$_{\rm H,av}=6.0_{-1.3}^{+**}\times10^{24}$\,cm$^{-2}$ are in agreement with the results found by \citet{Marchesi_2019}. The constants accounting for the intensity of the reprocessed emission component are $A_{S,90}=0.95_{-0.15}^{+0.21}$ and $A_{S,0}=0.32_{-0.07}^{+0.13}$, suggesting a forward-reflected dominated scenario, meaning that the majority of the photons scattered toward us pass through the medium between us and the torus. This can point either to an edge-on scenario or to a very large covering factor. 
These results are in accordance with those obtained using both \borus\ and \uxclumpy\ (see Section \ref{subsec:borus_results}, \ref{subsec:uxclumpy_results}).

\subsection{UXCLUMPY}\label{subsec:uxclumpy_results}
\begin{figure}[ht]
    \centering
    \includegraphics[trim={0cm 1cm 0 0cm},clip, width=0.5\textwidth]{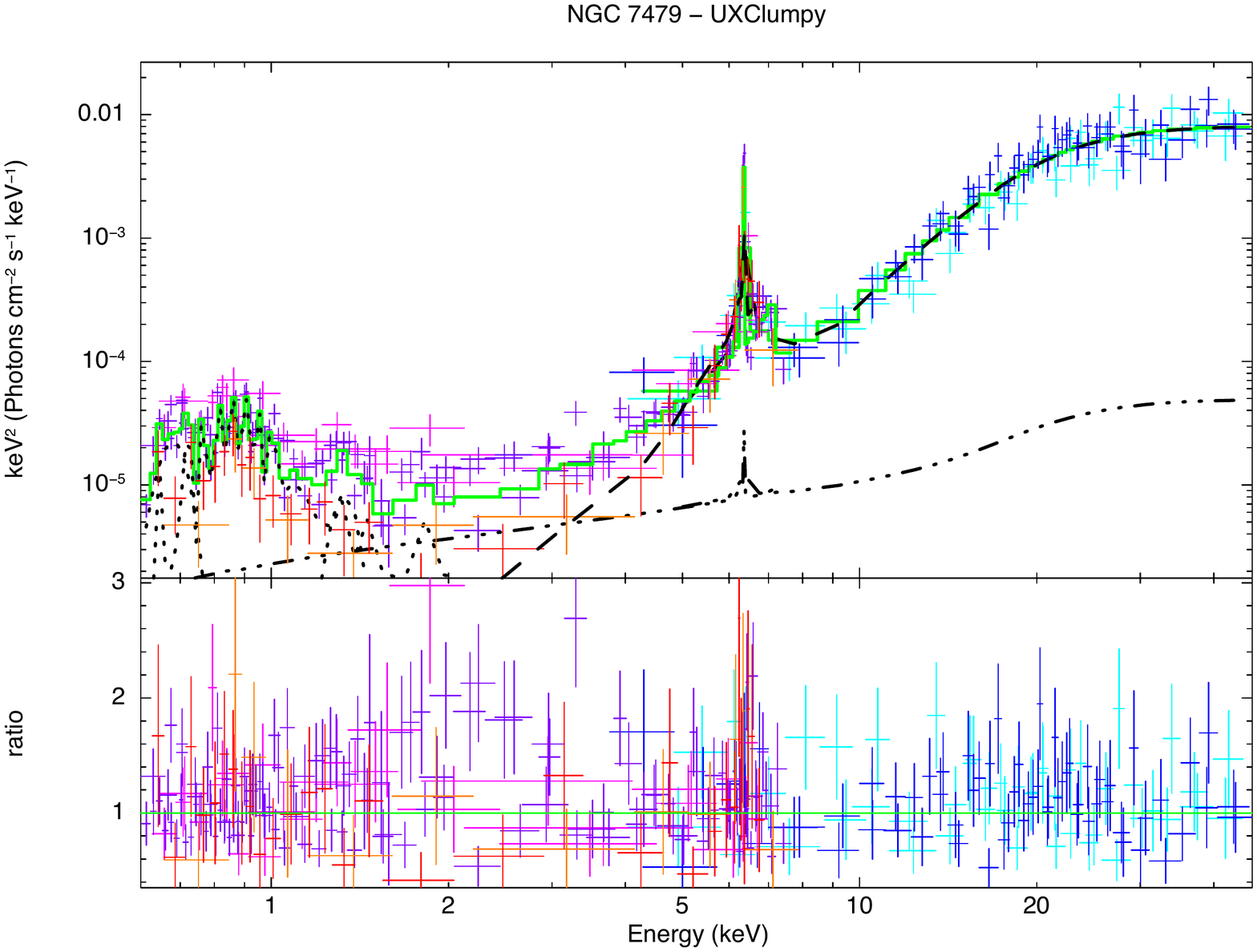}
    \caption{Unfolded \chandra\ (orange and red) \xmm\ (purple and magenta) and \nustar\ (blue and cyan) 0.6-50\,keV combined spectra of NGC 7479 modeled with \uxclumpy. The best-fit model is plotted with a solid green line, while the individual model components are plotted in black. Reflection component: dashed line while. Scattering component: dash-dotted.  \texttt{apec}:  dotted line.}
    \label{fig:uxclumpy_spectrum}
\end{figure}

The 0.6-50\,keV best-fit model is reported in Figure \ref{fig:uxclumpy_spectrum}.
As we were not able to constrain the inclination angle, in order to get consistent results with the previous models and to improve the goodness of the fit (from a reduced statistic of 1.24 to 1.22), we froze the inclination angle to be equal to the one we found using \borus, so $\theta_{i,\rm uxclumpy}=\theta_{i,\rm borus02}=37.8^{\circ}$.
The best-fit photon index is $\Gamma=1.52_{-0.05}^{+0.04}$; the best-fit value for \texttt{TORsigma}=$24.6_{-**}^{+20.1}$ and \texttt{CTKcover=}$0.60_{-0.07}^{+**}$ are compatible with a scenario in where the AGN has a thick inner ring of Compton-thick cloud in the central region surrounded by thinner clouds  dispersed in the outer region of the torus\footnote{An accurate visualization of this and other distributions of the obscuring material can be found at \url{https://github.com/JohannesBuchner/xars/blob/master/doc/uxclumpy.rst}}.

\begingroup
\renewcommand*{\arraystretch}{1.2}
\begin{table*}
\centering
\vspace{.1cm}
\caption{X-ray fitting results for NGC 7479}
  \begin{tabular}{ll|ccc}
  \textbf{Parameter} && \textbf{\borus}  & \textbf{\mytorus\ decoupled} & \textbf{\uxclumpy}\\  
            
 \hline\hline
     stat/d.o.f         &         & 476.9/390       & 480/390 &   463.9/382 \\ 
    
    red stat       &         &  1.22             & 1.23         & 1.21 \\
    \hline

\hline

  $kT$\footnote{\texttt{apec} model temperature in units of keV.}      &    & $0.79_{-0.03}^{+0.03}$      & $0.78_{-0.03}^{+0.03}$     & $0.80_{-0.02}^{+0.02}$      \\   
    
    $\Gamma$ \footnote{Powerlaw photon index.}     &     & $1.79_{-0.06}^{+0.01}$      & $1.69_{-0.10}^{+0.05}$    & $1.52_{-0.05}^{+0.04}$    \\   
 
    N$_{\rm H,av}$\footnote{Torus' average hydrogen column density in units of $10^{24}$\,cm$^{-2}$.} & & $17.3_{-11.7}^{+30.9}$        & $6.03_{-1.3}^{+**}$     & $\dots$\\

    C$_F$\footnote{Covering factor of the torus, as computed using \borus.}    &       & $0.88_{-0.01}^{+0.01}$      & $\dots$       & $\dots$  \\

    cos$(\theta_i)$\footnote{Cosine of the inclination angle, as computed using \borus. cos$(\theta_i)$=0 represents an edge-on scenario.}                &    & $0.79_{-0.01}^{+0.01}$      & $\dots$  & $\dots$        \\
    
    CTKcover\footnote{Covering factor of inner ring of clouds, as computed using \uxclumpy.} &  & $\dots$  & $\dots$ & $0.60_{-0.07}^{+**}$\\

    TOR$\sigma$\footnote{Cloud dispersion factor, as computed using \uxclumpy.} &  & $\dots$  & $\dots$ & $24.6_{-**}^{+20.1}$\\

    $A_{S,90}$\footnote{Reflection component constant associated with an edge-on scenario, as computed using \mytorus.}          &          &  $\dots$        & $0.95_{-0.15}^{+0.21}$   & $\dots$   \\

    $A_{S,0}$\footnote{Reflection component constant associated with a face-on scenario, as computed using \mytorus.}                    &  &  $\dots$   & $0.32_{-0.07}^{+0.13}$    & $\dots$  \\
    
     $F_s (\times 10^{-3} )$ \footnote{Scattering fraction.}      &   & $0.41_{-0.07}^{+0.07}$ & $0.54_{-0.23}^{+0.26}$ & $7.06_{-2.20}^{+2.80}$  \\
    
     Norm ($10^{-3}$)\footnote{Normalization of the intrinsic AGN emission.}       &&  $8.48_{-0.21}^{+0.87}$   & $6.07_{-0.16}^{+0.13}$  &$2.87_{-0.30}^{+0.30}$ \\
    \hline
     &\xmm\ -- a             & $1.00_{-0.12}^{+0.20}$         & $0.95_{-0.27}^{+0.29}$      & $0.98_{-0.44}^{+0.41}$\\
     
     &   \chandra\ -- a          &  $0.60_{-0.16}^{+0.19}$       & $0.61_{-0.30}^{+0.39}$    & $0.57_{-0.32}^{+0.38}$\\
      
   $C_{\rm AGN}$\footnote{Cross normalization constant between observations. $C_{\rm AGN}$ is fixed to 1 for \nustar\ observations.}  &   \chandra\ -- b         & $0.55_{-0.09}^{+0.11}$       & $0.56_{-0.17}^{+0.19}$   & $0.53_{-0.23}^{+0.27}$\\
    
& \nustar\ -- 1         & $1$        & $1$     & $1$         \\
    
     &   \xmm\ -- c             & $1.00_{-0.07}^{+0.03}$         & $0.83_{-0.11}^{+0.14}$      & $0.86_{-0.29}^{+0.28}$\\
     
    &   \nustar\ -- 2           & $1.00_{-0.05}^{+0.07}$         & $0.93_{-0.18}^{+0.22}$      & $0.88_{-0.30}^{+0.30}$\\

     \hline
     
         &  { \xmm\ -- a  }        & $-12.55_{-0.04}^{+0.04}$   & $-12.56_{-0.04}^{+0.04}$  & $-12.49_{-0.17}^{+0.11}$   \\
  
   & \chandra\ -- a       & $-12.57_{-0.11}^{+0.10}$  & $-12.57_{-0.11}^{+0.10}$  & $-12.56_{-0.31}^{+0.17}$ \\     
  
 Log(F$_{2,10}$)\footnote{Observed flux between 2-10\,keV computed using \texttt{cflux} command.}  & \chandra\ -- b     & $-12.52_{-0.06}^{+0.06}$   & $-12.52_{-0.06}^{+0.06}$ & $-12.51_{-0.11}^{+0.08}$\\ 
  
     & \nustar\ -- 1        & $-12.65_{-0.03}^{+0.03}$ & $-12.61_{-0.03}^{+0.03}$  & $-12.45_{-0.09}^{+0.08}$     \\
    
    & \xmm\ -- c           & $-12.63_{-0.01}^{+0.01}$  & $-12.61_{-0.01}^{+0.01}$ & $-12.60_{-0.04}^{+0.03}$\\
    
    &  \nustar\ -- 2       & $-12.64_{-0.03}^{+0.02}$  & $-12.62_{-0.03}^{+0.02}$ & $-12.47_{-0.06}^{+0.05}$   \\   

  \hline
 
  Log(F$_{10,40}$)\footnote{Observed flux between 10-40\,keV computed using \texttt{cflux} command.} & \nustar\ -- 1        & $-11.12_{-0.03}^{+0.03}$ & $-11.14_{-0.03}^{+0.03}$  & $-11.16_{-0.04}^{+0.03}$     \\
    
    &  \nustar\ -- 2       & $-11.06_{-0.03}^{+0.02}$  & $-11.07_{-0.03}^{+0.03}$  & $-11.11_{-0.03}^{+0.03}$  \\ 
    
  

    
    

  \hline
\end{tabular}
  \label{table:fitting}
\vspace{.2cm}
\end{table*}
\endgroup

\subsection{Line-of-sight hydrogen column density variability over time}\label{subsec:nh_var}
\begin{figure*}[ht]
    \centering
    \includegraphics[width=1\textwidth]{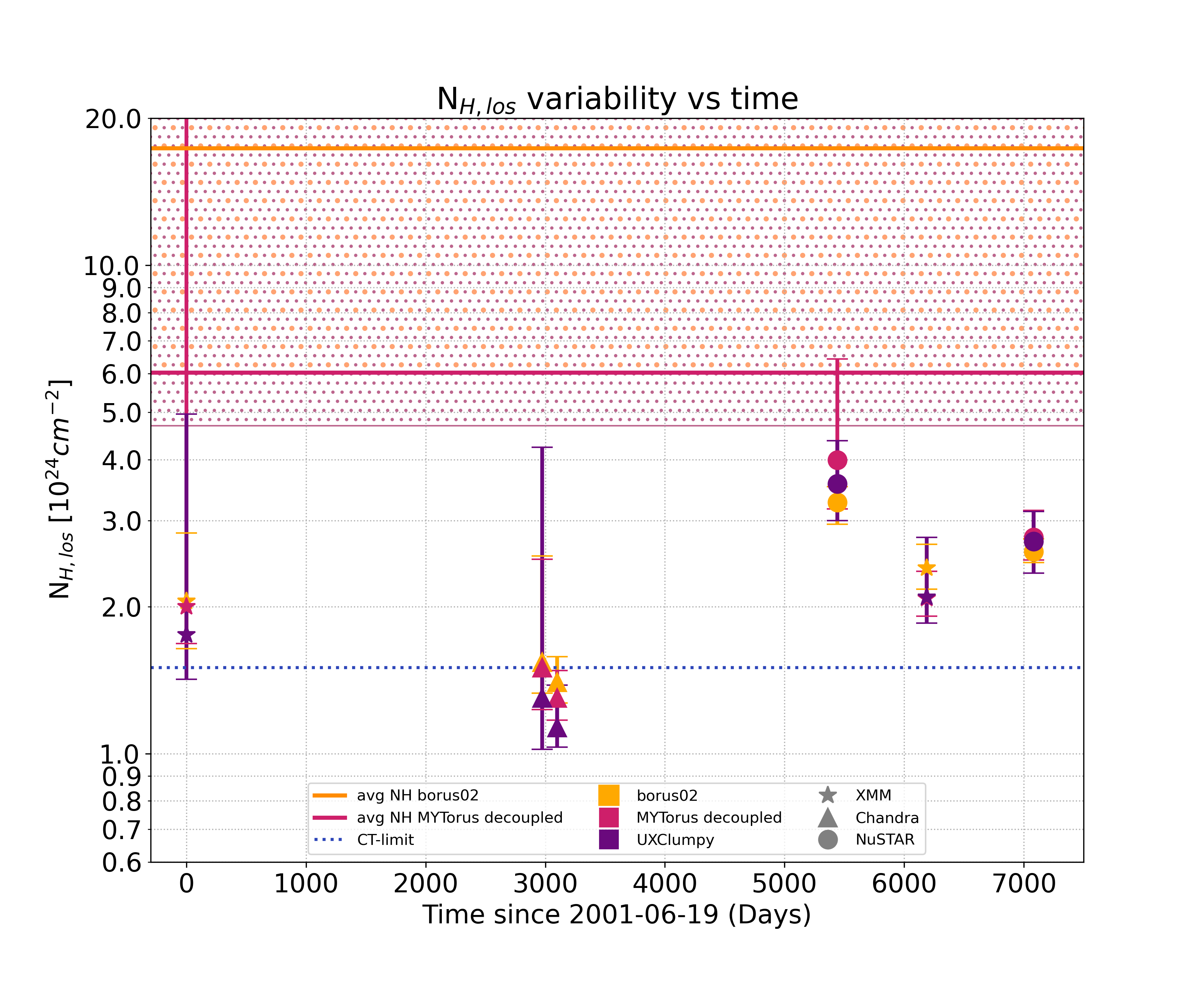}
    \caption{Line-of-sight hydrogen column density variability of NGC 7479 between 2001 and 2020. We note how the N$_{\rm H,av}$ of the torus is higher with respect to the N$_{\rm H,los}$ for all model.  For clarity purposes, we added 50 days to the date of the second \chandra\ observation (\chandra\ -- b in Table \ref{table:sources}).}
    \label{fig:nh_var}
\end{figure*}

In Table \ref{table:nh_values} we report line-of-sight hydrogen best-fit values, as a result of our multi-epoch X-ray spectral analysis of NGC 7479; the variability can be visualized in Figure \ref{fig:nh_var}.
We note that, despite the models being able to provide errors (i.e. constraints) for N$_{\rm H,los}$ in almost all cases, confidence contours (see Figure \ref{fig:cont_plots} -- \textit{left}, in Appendix \ref{appendix:cont_plots}) show that the degeneracy between N$_{\rm H,los}$ and C$_{\rm AGN}$ is not fully broken in all scenarios. In fact, for \chandra\ -- a observation we are unable to place upper limits for N$_{\rm H,los}$, based on the contours. By contrast, the degeneracy between N$_{\rm H,los}$ -- cos($\theta_i$) and N$_{\rm H,los}$ -- $\Gamma$ is fully broken for all observations within 1$\sigma$ as shown in Figure \ref{fig:cont_nh_costheta_gamma} in Appendix \ref{appendix:cont_plots}. \footnote{Considering the good agreement between the three models used in this work, confidence contour plots have been computed using only the \borus\ model.} 


\begingroup
\renewcommand*{\arraystretch}{1.2}
\begin{table*}
\centering
\vspace{.1cm}
\caption{Line-of-sight hydrogen column density best-fit values for NGC 7479 in units of $10^{24}$\,cm$^{-2}$}
  \begin{tabular}{lc|ccc}
  \textbf{Observation} & \textbf{Date}& \textbf{\borus}  & \textbf{\mytorus\ decoupled} & \textbf{\uxclumpy}\\ 
     \hline
      \xmm\ -- a &  2001-06-19        & $2.05_{-0.41}^{+0.78}$   & $2.00_{-0.32}^{+**}$  & $1.75_{-0.33}^{+3.21}$   \\
    
     \chandra\ -- a & 2009-08-11     & $1.54_{-0.21}^{+1.00}$  & $1.50_{-0.27}^{+1.00}$  & $1.30_{-0.28}^{+2.94}$ \\     
  
  \chandra\-- b & 2009-10-24    & $1.40_{-0.13}^{+0.18}$   & $1.30_{-0.13}^{+0.18}$ & $1.13_{-0.10}^{+0.25}$\\

      \nustar\ -- 1 & 2016-05-12     & $3.27_{-0.32}^{+0.25}$ & $3.99_{-0.82}^{+2.44}$ & $3.57_{-0.57}^{+0.80}$      \\
    
     \xmm\ -- c  & 2018-05-30        & $2.40_{-0.23}^{+0.28}$  & $2.08_{-0.17}^{+0.28}$ & $2.09_{-0.24}^{+0.68}$\\    
    
      \nustar\ --2 & 2020-11-06     & $2.59_{-0.13}^{+0.16}$  & $2.77_{-0.28}^{+0.38}$ & $2.72_{-0.38}^{+0.41}$    \\   
       
  \hline
\end{tabular}
  \label{table:nh_values}
\vspace{.2cm}
\end{table*}
\endgroup

\begin{figure}[ht]
    \centering
    \includegraphics[trim={0.5cm 0cm 0 0cm},clip, width=0.47\textwidth]{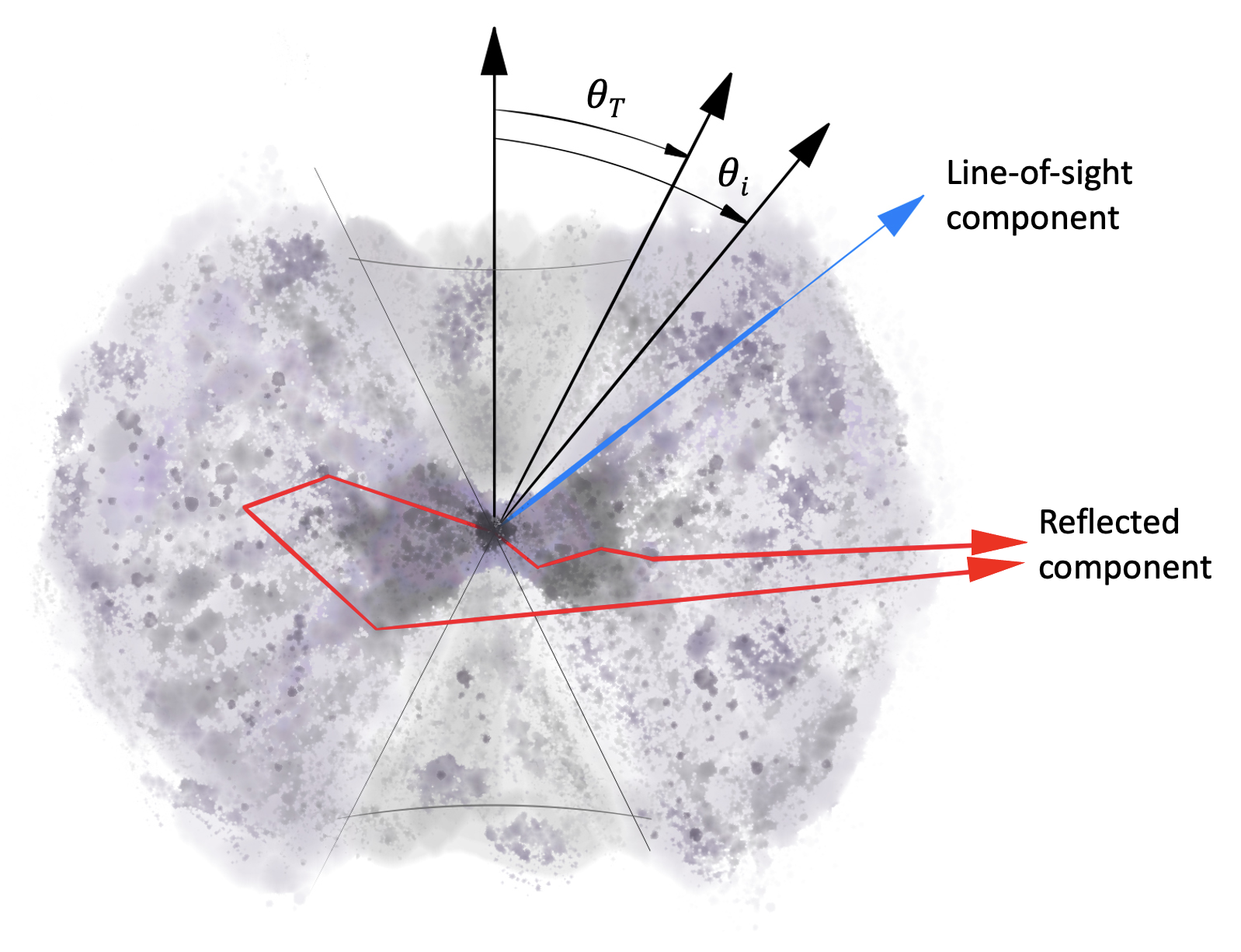}
    \caption{NGC 7479's torus representation based on best-fit parameters obtained during the modeling.}
    \label{fig:torus_cartoon}
\end{figure}

Interestingly, we notice how the average column density of the torus always remains above the line-of-sight values for both \mytorus\ and \borus, suggesting a reflection-dominated scenario. We note that the average hydrogen column density of the torus is disentangled from the inclination angle, as suggested by the confidence contour plot in Figure \ref{fig:cont_plot_costheta_c2_nhav_gamma}, \textit{bottom right}.This is in agreement with the high covering factor computed by \borus\ ($C_F=0.88^{+0.01}_{-0.01}$) and the best-fit values found with \uxclumpy, for which the dispersion of the clouds is \texttt{TOR}$\sigma=24.6_{-**}^{+20.1}$ and the covering factor of the inner reflecting material is \texttt{CTKcover}$=0.60_{-0.07}^{+**}$. This suggests the presence of an inner ring of dense gas \citep[][]{Buchner_2019, honig2019} surrounded by a patchy-clumpy torus made of dispersed low-density clouds.
This result is in agreement with those obtained by \citet{Laha2020}, in where the addition of a partial-cover absorber component is required to model 11/20 of the sources analyzed in their work. 
A schematic representation of the torus can be seen in Figure \ref{fig:torus_cartoon}. In this case, we used the best-fit data obtained from our analysis (such as covering factor, cloud distribution, and inclination angle) to model how the torus of NGC 7479 would look like, according to our results. This leads us to consider a scenario in which the reflector (i.e. inner ring) and the absorber (i.e. cloud distribution) are two different structures within the torus; rather than the whole structure being responsible for both emissions (as often assumed). 
In this scenario, the inner dense Compton-thick ring generates the reflection-dominated spectra we see in Figures \ref{fig:borus_spectrum}, \ref{fig:myt_spectrum}, and \ref{fig:uxclumpy_spectrum}; while the more dispersed clouds, with variable densities, are responsible for the subdominant line-of-sight component. As it can be seen in Figure \ref{fig:nh_var}, the variability trend is shown by all the three models, indicating that, despite the reflection-dominated scenario, the quality of the data allows us to confirm a change in the line-of-sight column density.

Using the data we obtained from our N$_{\rm H,los}$ analysis, we compute the upper limit of the cloud distance from the central engine (R), following the approach described in \citet{risaliti_2005}. In here, the distance $R\sim600t_{\rm 100}^2n_{10}^2\Delta N_{\rm H,24}^{-2}R_{\rm S}$ in where $t_{\rm 100}$ is the variability time in units of 100\,ks, $\Delta$N$_{\rm H,24}$ is the l.o.s. column density difference between two consequent observations, in units of $10^{24}$\,cm$^{-2}$ (we can assume it to be the density of the cloud obscuring the central engine during that time); and $R_{\rm S}$ is the Schwarzschild radius. The cloud density ($n_{10}$) in units of $10^{10}$\,cm$^{-3}$ can be computed by simply dividing the $\Delta$N$_{\rm H}$ by the size of the corona, shown to be comprised between 3$R_{\rm S}$ and 20 $R_{\rm S}$ \citep[][]{risaliti_2005, fabian_2015}. Using the correlation between stellar velocity dispersion and black hole mass found by \citet{tremaine_2002}\footnote{The intrinsic scatter of the $M_{\rm BH}-\sigma$ relation considered is $\epsilon \sim$ 0.25-0.30 dex, as reported in \cite{tremaine_2002}.} and the central stellar velocity dispersion reported by \citet{mcelroy_1995}, we are able to estimate the mass of the central engine to be $log(M_{BH}/M_\odot)= 7.07$, which is in accordance with the mass computed by \citet{panessa2006} $log(M_{BH}/M_\odot)= 7.09$, and from here estimate $R_{\rm S,NGC7479}\simeq3.87\times 10^{12}$\,cm. The results of our calculations are shown in Table \ref{table:distance}, in where the clouds distance is reported only when the N$_{\rm H,los}$ between two consecutive observations is not compatible within errors.
\begin{table*}
\centering
\caption{Clouds properties}
   \begin{tabular}{l|cccc}
\hline
                             & &  & \textbf{D$\sim$3R$_{\rm S}$}  & \textbf{D$\sim$20R$_{\rm S}$}\\
        \textbf{Observations} & \textbf{$\Delta NH$}       & \textbf{$\Delta t$}& R    &  R   \\
                          &   [$\times 10^{24}$\,cm$^{-2}$] &      [100 ks]   &  [pc]      &    [pc]  \\
        \hline     
        \hline
        \chandra\ -- b | \nustar\ -- 1 & 1.87 & $2.07\times10^3$ & $2.37\times 10^{5}$ &     $5.34\times 10^{3}$    \\
        
        \nustar\ -- 1 | \xmm\ -- c & 0.87 & $6.64\times10^2$&  $2.31\times 10^{4}$ &           $5.20\times 10^{2}$\\
        \hline
        \end{tabular}
        \label{table:distance}
        
\end{table*}

Given the general good agreement between the three models used, the values we obtain do not change significantly when each model N$_{\rm H,los}$ is used; for simplicity the distances we report are calculated considering \borus\ N$_{\rm H,los}$ values.  What we obtain is an upper limit of the cloud distance, due to the large time window between the observations, in where multiple eclipsing events may have happened. Indeed, only observations over time-scales of days to  weeks to months can give a more accurate analysis and vision of the cloud dynamics of NGC 7479 and allow us to better constraint the properties of the obscuration region. Such observations would allow us to better understand the physical differences between the absorber and the reflector, in terms of position and density, as well as their scale. 



\subsection{Comparison with previous results}
NGC 7479 was found to be a X-ray CT-AGN candidate by \citet{Marchesi_2019} and \citet{zhao_2021} by using \xmm\ (ObsID 0824450601) and \nustar\ (ObsID 60201037002)  observations fitted with the \mytorus\ model and \borus\ in the following configuration ($C_F=0.5$, $\theta_i=87.7^{\circ}$). Although both cases showed the same goodness of fit, they had some discrepancy in the results. Our large dataset is able to break the degeneracy model and show that a solution like the one found in \citet{Marchesi_2019} is preferred, in where our photon index and average $N_{\rm H}$ are in perfect agreement within errors. 
Our best-fit values are also in accordance, within errors, with the ones reported by \cite{tanimoto2022}, in where one \nustar\ and one \xmm\ observations, as well as \swift/XRT data are fitted using the \texttt{XClumpy} code \citep[]{Tanimoto2019}. They find $\Gamma=1.88_{-0.16}^{+0.12}$ and an average hydrogen column density of N$_{\rm H,av}=15.3_{-3.25}^{+13.5}\times 10^{24}$\,cm$^{-2}$, both in agreement with our results.

\subsection{Bolometric luminosity, SMBH mass and Eddington ratio}
We report the 2-10\,keV, 10-40\,keV and 0.6-300\,keV intrinsic luminosity of each observation in Table \ref{table:luminosities} in the Appendix. From here, applying the bolometric correction reported in \citet{vasudevan2010}, we are able to compute the bolometric luminosity of the source, which represents the measurement of the total AGN emission over all electromagnetic energies. Over all epochs, we get a range of $L_{\rm bol}=4.48-8.16\times 10^{43}$\,erg\,s$^{-1}$.\footnote{For simplicity, we are using the values from our \borus\ best-fit.} We can now derive the Eddington luminosity $L_{\rm Edd}=\frac{4\pi G M_{\rm BH}m_p c}{\sigma_T}$, where $m_p$ is the mass of the proton, $M_{\rm BH}$ is the mass of the black hole and $\sigma_T$ is the Thompson cross section. Considering the mass we just found, $L_{\rm Edd}=1.47\times 10^{45}$\,erg\,s$^{-1}$. This leads us to the computation of an Eddington ratio range of $\lambda_{\rm Edd}=0.03-0.05$, where  $\lambda_{\rm Edd}=L_{\rm Bol}/L_{\rm Edd}$. This result is in agreement the value found by \cite{tanimoto2022} of $\lambda_{\rm Edd}=0.037$.

The bolometric luminosity can be also derived by adding to the total X-ray luminosity\footnote{We measured the luminosity over all the energy ranges taken into account in our best-fit.}  the far-infrared luminosity of the source \citep[][]{Lusso_2012}. From \citet{soifer_2004} we get that $L_{\rm FIR}=6.35 \times 10^{43}$\,erg\,s$^{-1}$. Adding this to our total X-ray luminosity we get a range of $L_{\rm bol}=6.9-7.5 \times 10^{43}$\,erg\,s$^{-1}$, which corresponds to an Eddington ratio range of $\lambda_{\rm Edd}=0.04-0.05$, in perfect accordance with the previous result we obtained.

\section{Conclusions}\label{sec:conclusion}

In this paper we presented the joint fitting of \chandra, \xmm\ and \nustar\   observations of NGC 7479, a nearby CT-AGN, using the \borus, \mytorus, and \uxclumpy\ models, leading to the following conclusions:
\begin{enumerate}
    \item The simultaneous multi-epoch analysis of NGC 7479 X-ray spectra allowed us to put tight constrains on its torus global properties such the covering factor, the inclination angle and the average column density.  
    \item We validate the clumpy-torus scenario by measuring the line-of-sight hydrogen column density variability over a time span of $\sim$20 years. 
    Thanks to such variability, we were able to put an upper limit on the distance of the clouds to the central X-ray emitter.
    \item The best-fit parameters suggest the disentanglement of the reflection and the absorption material inside the torus. In particular, the presence of a thick \textit{inner ring} of reflecting medium is required to explain the reflection-dominated spectra.
    \item Starting from the X-ray $L_{2-10}$ luminosity, we calculate the bolometric luminosity of $L_{\rm bol}=4.48-8.16\times 10^{43}$\,erg\,s$^{-1}$ and the Eddington ratio $\lambda_{\rm Edd}=0.03-0.05$ over all epochs for NGC 7479. These are in agreement with what is found in the literature.
\end{enumerate}

\section*{acknowledgements}

A.P., N.T., R.S. and M.A. acknowledge funding from NASA and SAO  (contracts 80NSSC19K0531, 80NSSC20K0045, 80NSSC21K0016 and GO0-21083X). S.M. acknowledges funding from the the INAF ''progetti di Ricerca di Rilevante Interesse Nazionale'' (PRIN), Bando 2019 (project: ''Piercing through the clouds: a multiwavelength study of obscured accretion in nearby supermassive black holes''. The
scientific results reported in this article are based on observations
made by the X-ray observatories NuSTAR, XMM-Newton and Chandra. This research has made use of the NuSTAR Data Analysis Software (NuSTARDAS) jointly developed by the ASI Space Science Data Center (SSDC, Italy) and the California Institute of Technology (Caltech, USA). We acknowledge the use of the software packages XMM-SAS and HEASoft.

\bibliographystyle{aasjournal}
\bibliography{bibliography}

 \appendix
    
    \section{X-ray fitting results for NGC 7479}\label{sec:allbest}

\begin{table*}[ht]
\centering
\vspace{.1cm}
\caption{X-ray fitting results for NGC 7479}
\resizebox{\textwidth}{!}{%
  \begin{tabular}{ll|ccc}
  \textbf{Parameter} && \textbf{\borus}  & \textbf{\mytorus\ decoupled} & \textbf{\uxclumpy}\\  
            
 \hline\hline
    stat/d.o.f         &         & 476.9/390       & 480/390 &   463.9/382 \\ 
    
    red stat       &         &  1.22             & 1.23         & 1.21 \\
    \hline
    
       $kT$\footnote{\texttt{apec} model temperature in units of keV.}      &    & $0.79_{-0.03}^{+0.03}$      & $0.78_{-0.03}^{+0.03}$     & $0.80_{-0.02}^{+0.02}$  \\ 
   \hline 
   &  \xmm\ -- a        & $0.99_{-0.12}^{+0.12}$    &$0.27_{-0.06}^{+0.07}$    &$13.60_{-2.57}^{+2.78}\times 10^{-3}$      \\
    
     &  \chandra\ -- a     & $0.18_{-0.08}^{+0.10}$    &$0.05_{-0.02}^{+0.03}$    &$2.54_{-1.27}^{+1.61}\times 10^{-3}$         \\
    
  $C_{\rm apec}$\footnote{\texttt{apec} constant. $C_{\rm apec}$ is fixed to one for \nustar\ observations.}     & \chandra\ -- b       &$0.31_{-0.06}^{+0.07}$      &$0.09_{-0.02}^{+0.03}$    &$4.27_{-1.09}^{+1.29}\times 10^{-3}$     \\
   
    &  \nustar\ -- 1    & $1$              & $1$              & $1$                
\\

  & \xmm\ -- c       & $0.66_{-0.03}^{+0.03}$    &$0.19_{-0.03}^{+0.04}$    &$9.19_{-1.59}^{+1.47}\times 10^{-3}$        \\
    
    &   \nustar\-- 2   & $1$              & $1$              & $1$                              \\

    \hline
      Line 1 - Energy     &      & $0.68_{-0.01}^{+0.01}$      & $0.69_{-0.01}^{+0.01}$      & $0.68_{-0.02}^{+0.01}$         \\
      
      Line 1 - $\sigma$    &       & $0.03_{-0.02}^{+0.02}$      & $0.01_{-**}^{+0.02}$     & $0.04_{-0.02}^{+0.05}$       \\
      
      Line 1 - norm$(10^{-5})$      &    & $0.84_{-0.18}^{+0.54}$      & $2.25_{-0.65}^{+1.15}$    & $66.23_{-18.17}^{+52.69}$      \\

\hline
      Line 2 - Energy      &     & $1.33_{-0.03}^{+0.03}$      & $1.33_{-0.03}^{+0.03}$     & $1.33_{-0.03}^{+0.03}$       \\
      
      Line 2 - $\sigma$     &         & $0.07_{-0.03}^{+0.04}$       & $0.03_{-0.03}^{+0.03}$       & $0.06_{-0.04}^{+0.04}$       \\
      
      Line 2 - norm$(10^{-5})$ & & $0.14_{-0.05}^{+0.05}$       & $0.48_{-0.16}^{+0.20}$         & $9.19_{-2.93}^{+3.28}$   \\

 \hline   



    



    
    
    
     \hline

    
  
  
  
  
  \hline

  
  
  
    
    

 
    
  
\end{tabular}}

  \label{table:allbestfit}
\vspace{.2cm}
\end{table*}

    \section{Intrinsic luminosity of NGC 7479}\label{sec:intrinsiclum}

\begin{table*}[ht]
\centering
\vspace{.1cm}
\caption{Intrinsic luminosity of NGC 7479 in units of $10^{42}$\,erg\,s$^{-1}$}

  \begin{tabular}{c|l|ccc}
  \textbf{Parameter} & \textbf{Observation}&\textbf{\borus}  & \textbf{\mytorus\ decoupled} & \textbf{\uxclumpy}\\  
            
 \hline\hline

\hline
 &  \xmm\ -- a              & $4.08_{-0.49}^{+0.81}$      & $3.27_{-0.93}^{+1.00}$      & $2.09_{-0.94}^{+0.89}$      \\
     
      &  \chandra\ -- a          & $2.44_{-0.77}^{+0.65}$      & $2.10_{-1.03}^{+1.34}$      & $1.21_{-0.68}^{+0.81}$    \\
      
           $L_{2-10}$        &   \chandra\ -- b         & $2.24_{-0.45}^{+0.36}$      & $1.89_{-0.59}^{+0.65}$      & $1.13_{-0.49}^{+0.57}$   \\
    
                & \nustar\ -- 1             & $4.08$                      & $3.45$                  & $2.13$\\
    
     &   \xmm\ -- c               & $4.08_{-0.28}^{+0.12}$      & $2.86_{-0.38}^{+0.48}$      & $1.83_{-0.62}^{+0.60}$      \\
    
                &   \nustar\ -- 2           & $4.08_{-0.20}^{+0.28}$      & $3.20_{-0.62}^{+0.75}$      & $1.87_{-0.64}^{+0.64}$     \\
            
  \hline
   $L_{10-40}$     & \nustar\ -- 1             & $4.55$                      & $4.73$                     & $3.53$               \\
    
               & \nustar\ -- 2            & $4.55_{-0.22}^{+0.32}$      & $4.40_{-0.85}^{+1.04}$      & $3.10_{-1.06}^{+1.06}$          \\

  \hline
    

     
    
      
\end{tabular}
  \label{table:luminosities}

\end{table*}

\clearpage
 \section{Single epochs observations of NGC 7479}\label{appendix:single_ep}
\begin{figure}[ht]
    \centering
    \includegraphics[ width=0.4\textwidth ]{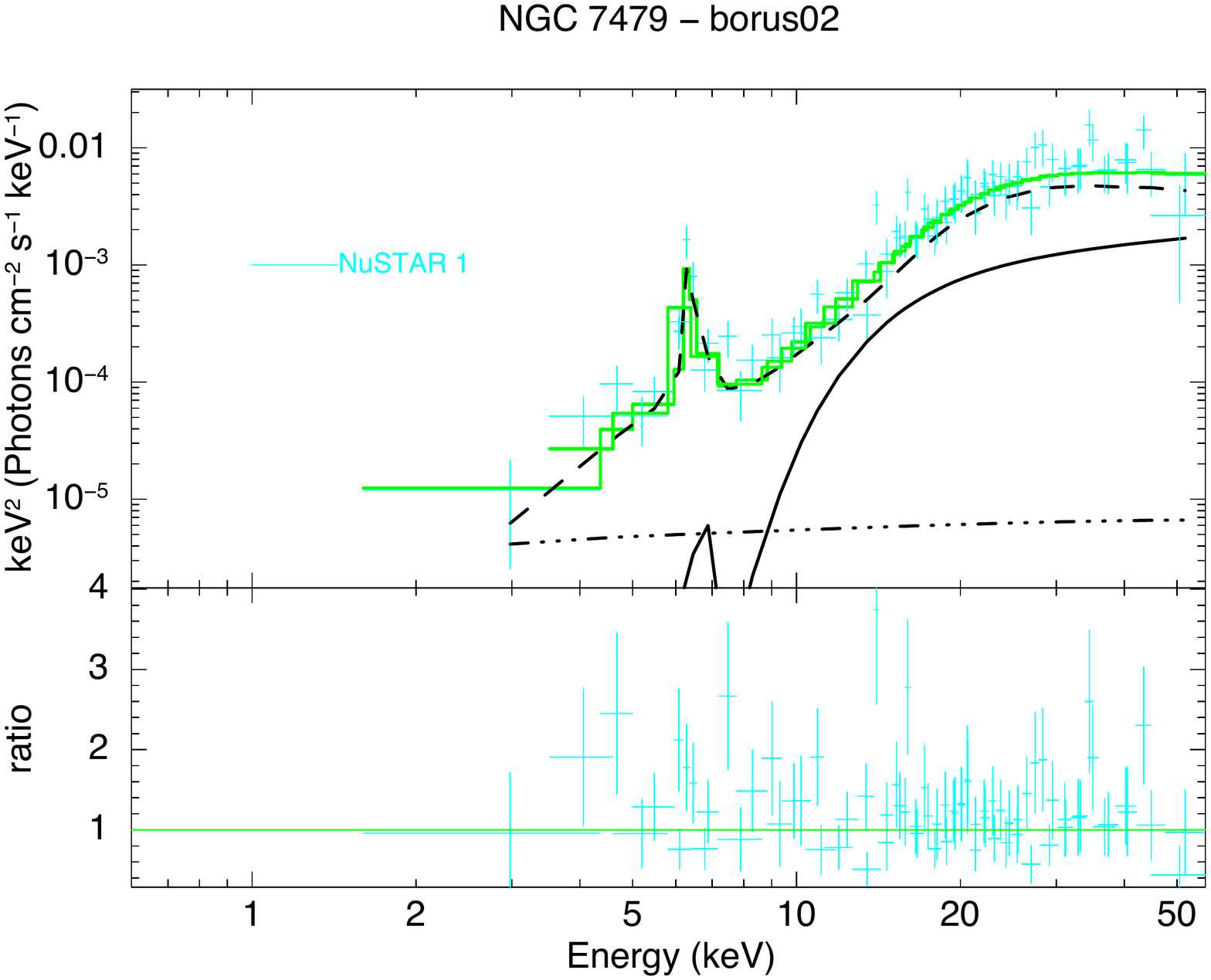}
    \includegraphics[ width=0.4\textwidth ]{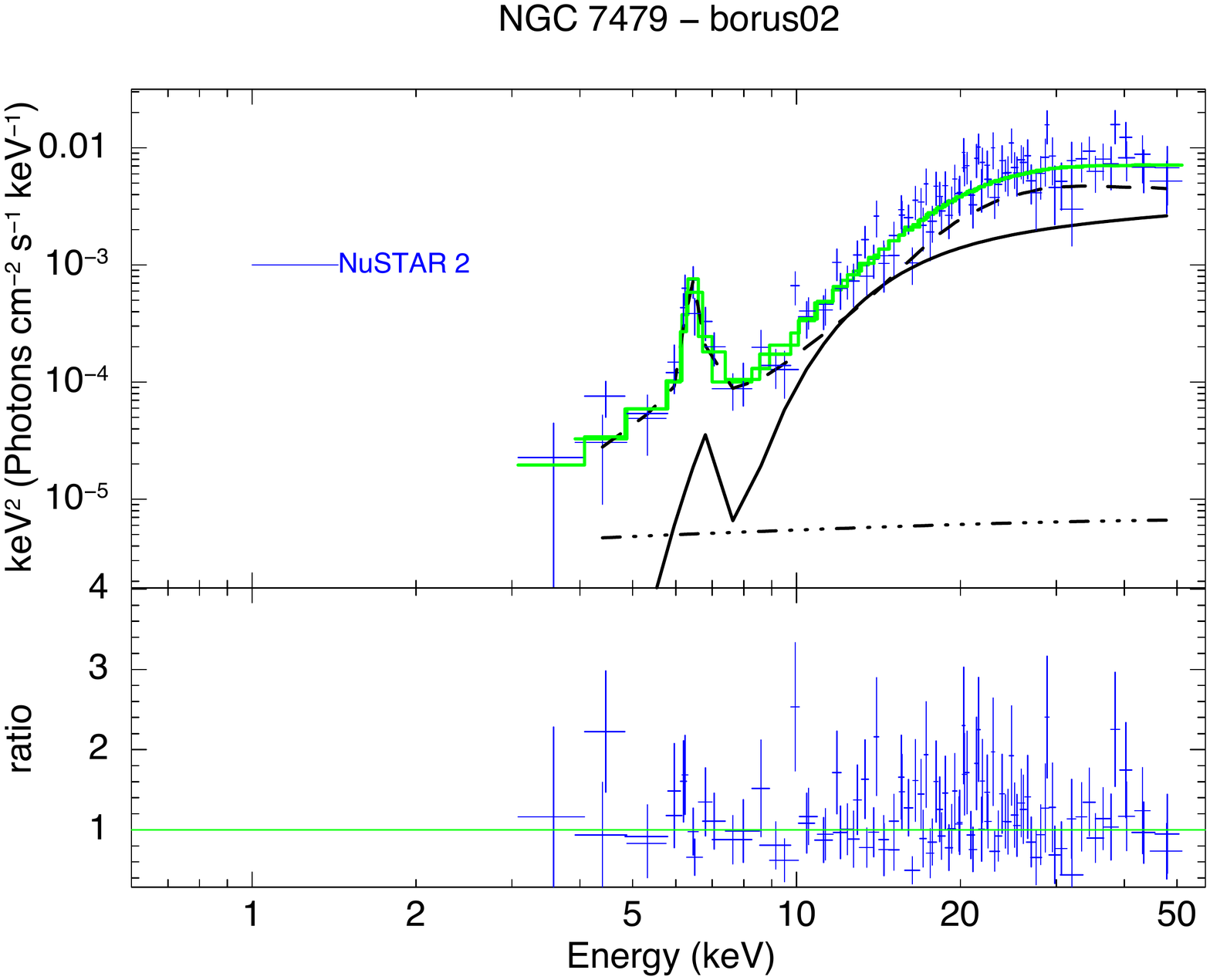}
    \includegraphics[ width=0.4\textwidth ]{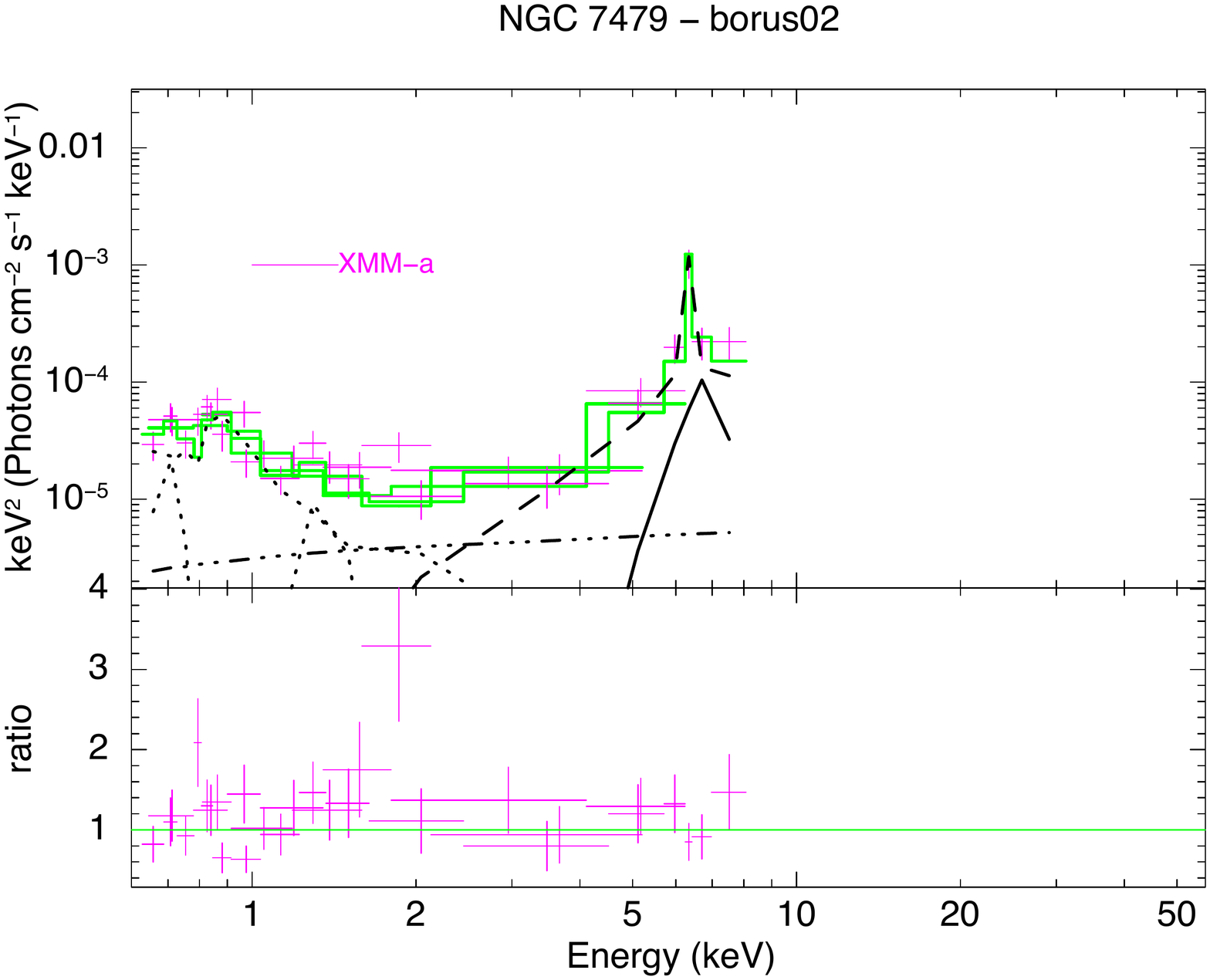}
    \includegraphics[ width=0.4\textwidth ]{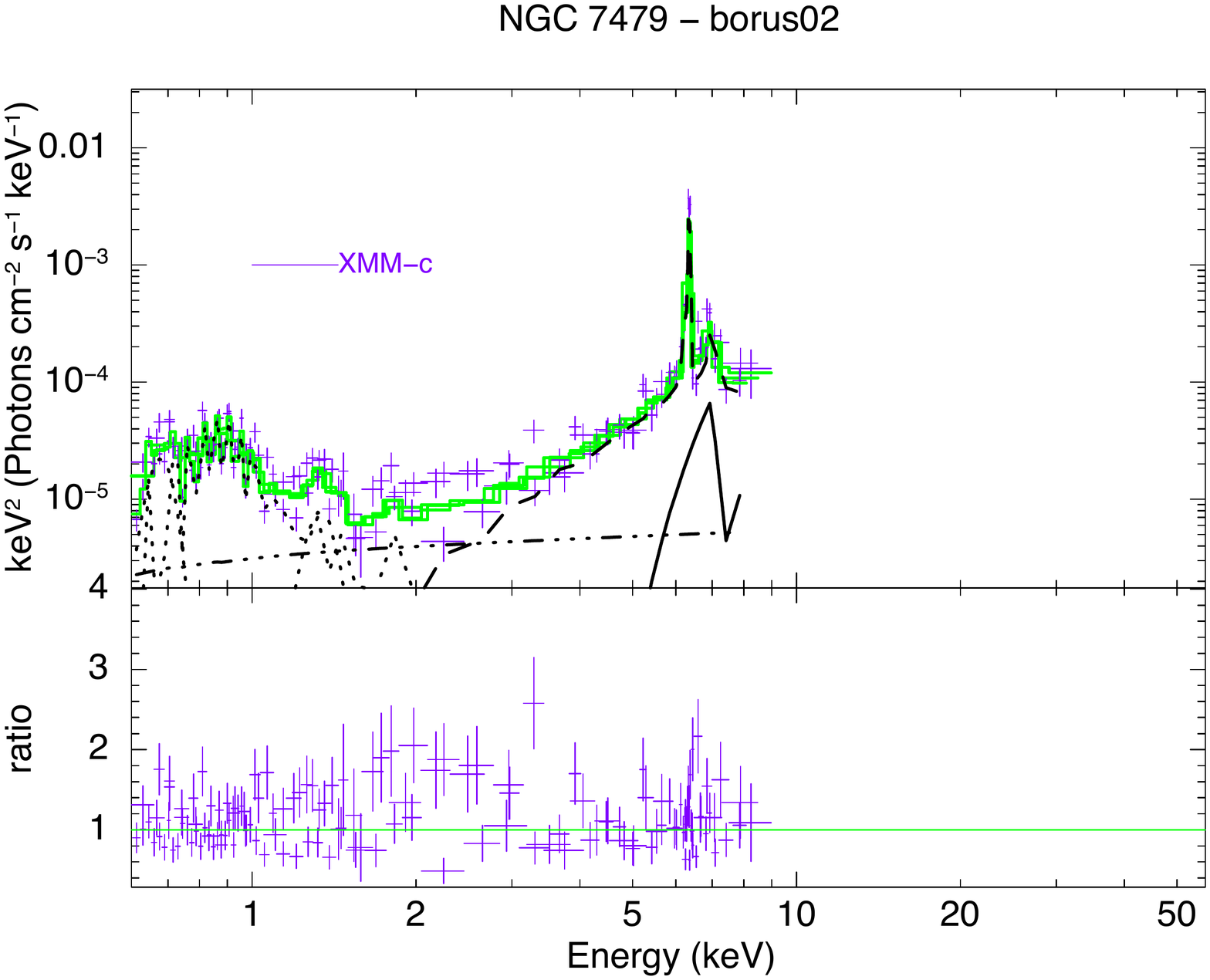}
    \includegraphics[ width=0.4\textwidth ]{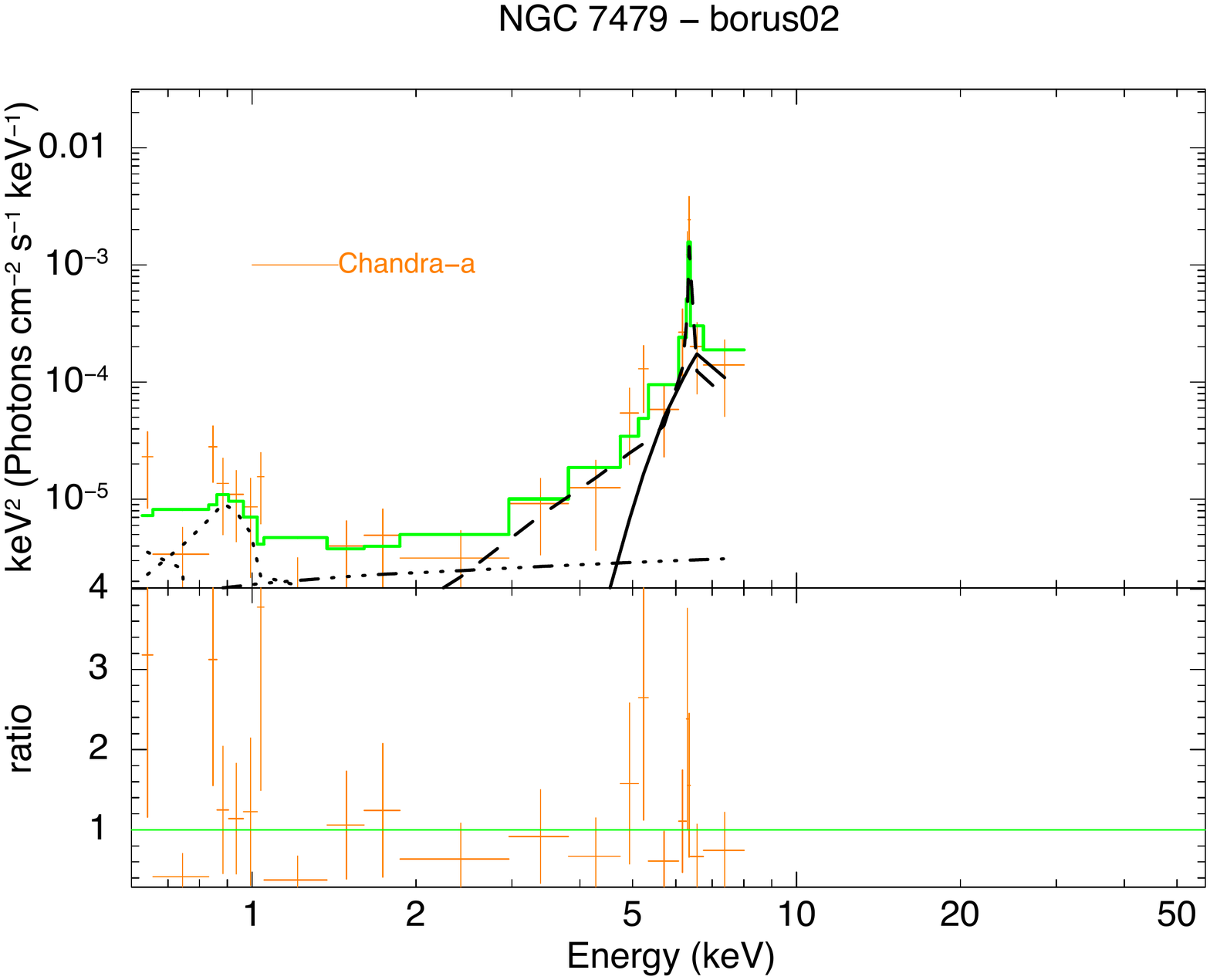}
    \includegraphics[width=0.4\textwidth ]{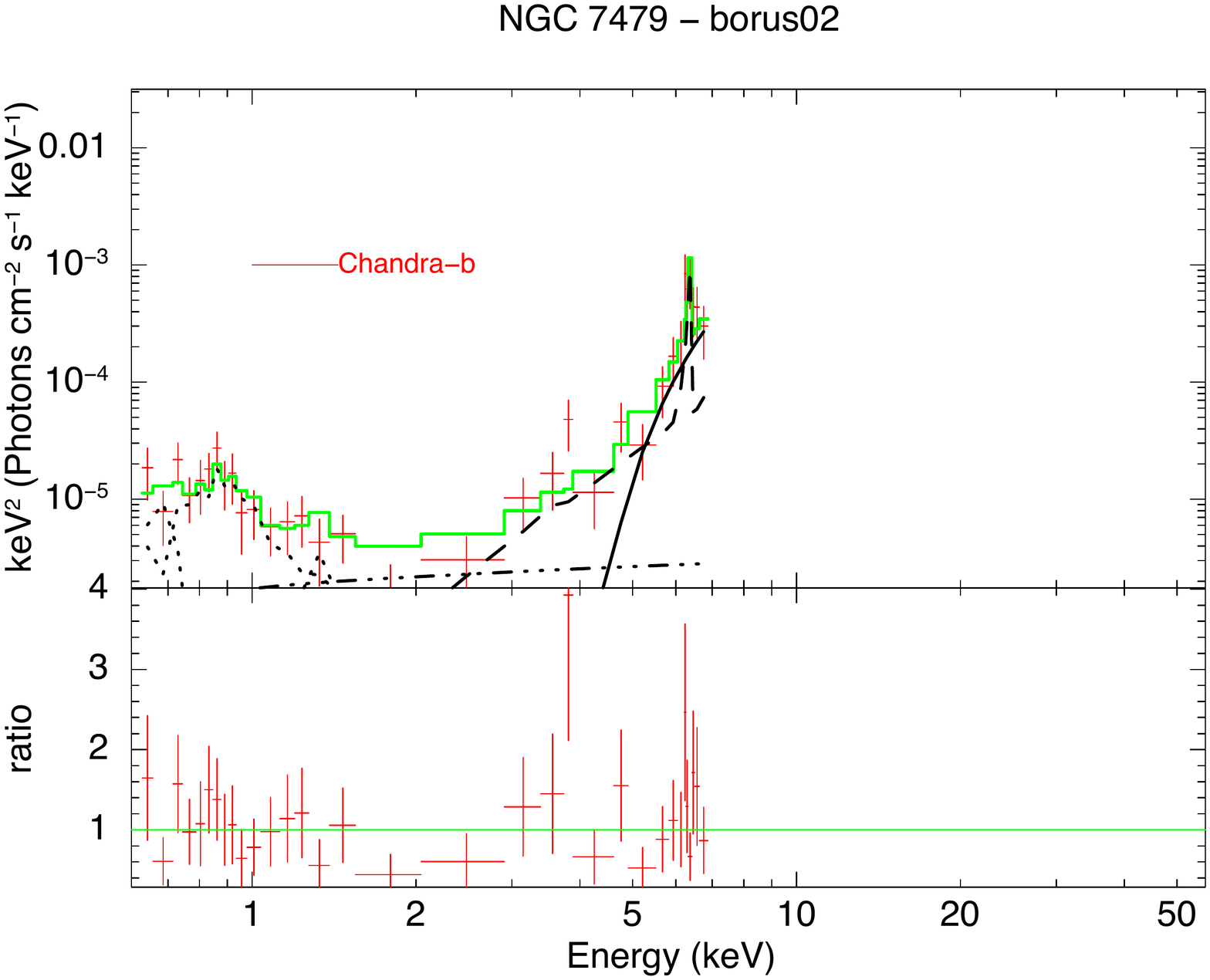}
    \caption{Unfolded \nustar\ (blue and cyan), \xmm\ (purple and magenta) and  \chandra\ (orange and red) 0.6-50\,keV single epochs spectra of NGC 7479 modeled with \borus. The best-fit model is plotted with a solid green line, while the individual model components are plotted in black. Line-of-sight component: solid line. Reflection component: dashed line. Scattering component: dash-dotted. \texttt{apec}: dotted line.}
    \label{fig:single_epochs}
\end{figure}

\section{Multi-dimensional contour plots}\label{appendix:cont_plots}
\begin{figure}[ht]
    \centering
    \includegraphics[ width=0.45\textwidth]{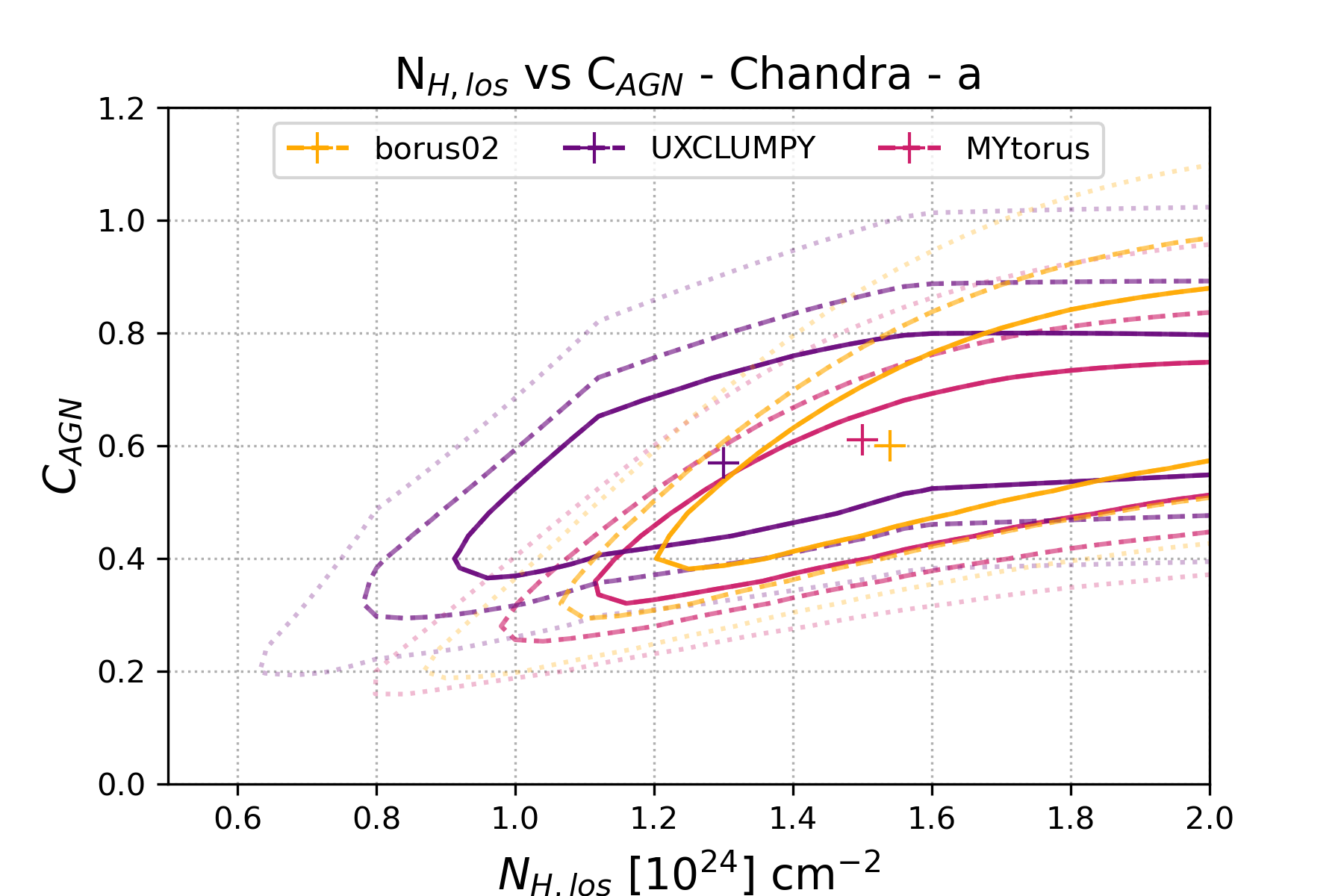}
    \includegraphics[ width=0.45\textwidth]{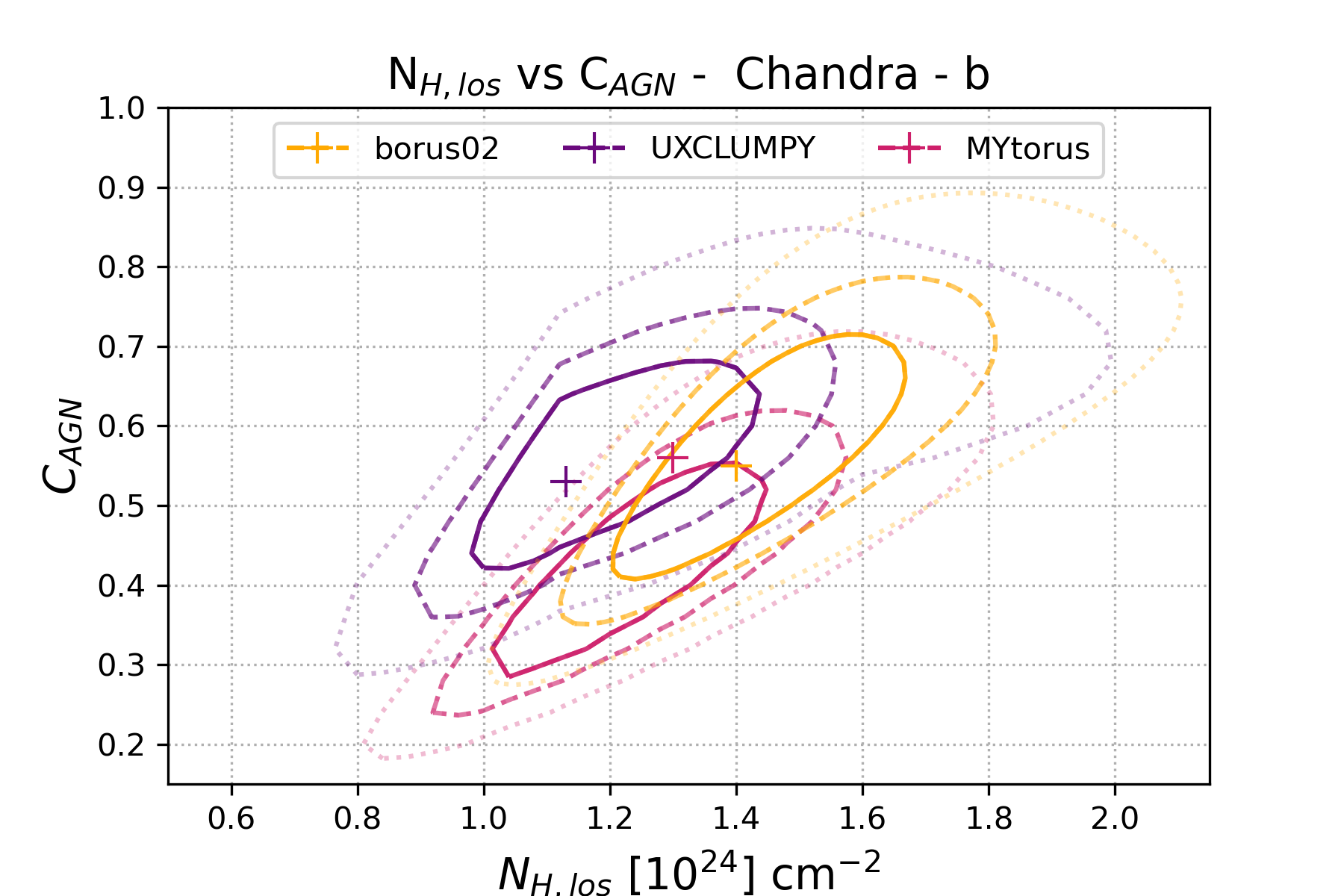}
    \caption{N$_{\rm H,los}$ -- C$_2$ multi-dimensional confidence contour plots for \chandra\ -- a (\textit{left}), and \chandra\ -- b observation (\textit{right}) obtained with the three physically motivated models used in this work. }
    \label{fig:cont_plots}
\end{figure}

\begin{figure}[ht]
    \centering
    \includegraphics[ width=0.45\textwidth]{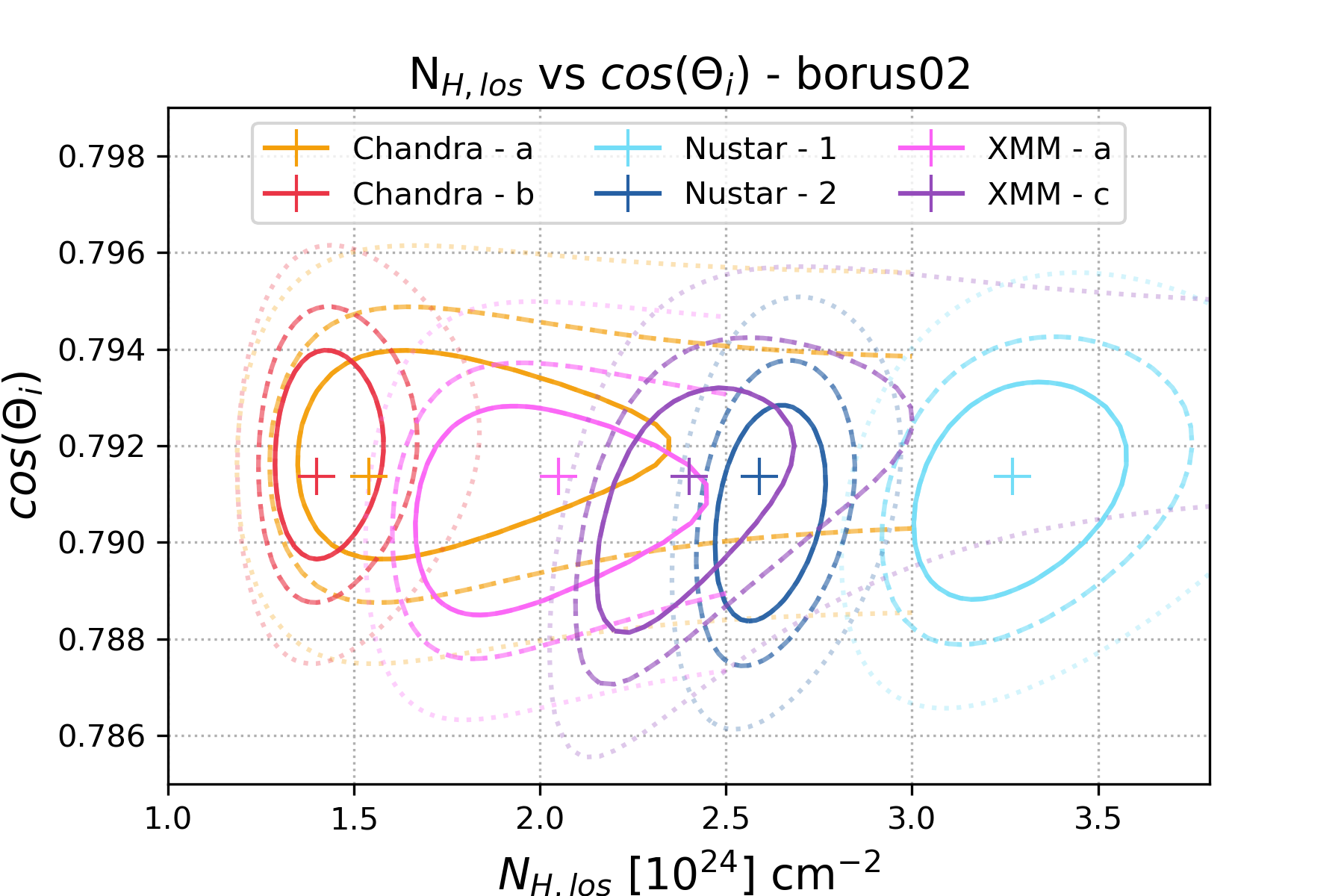}
    \includegraphics[ width=0.45\textwidth]{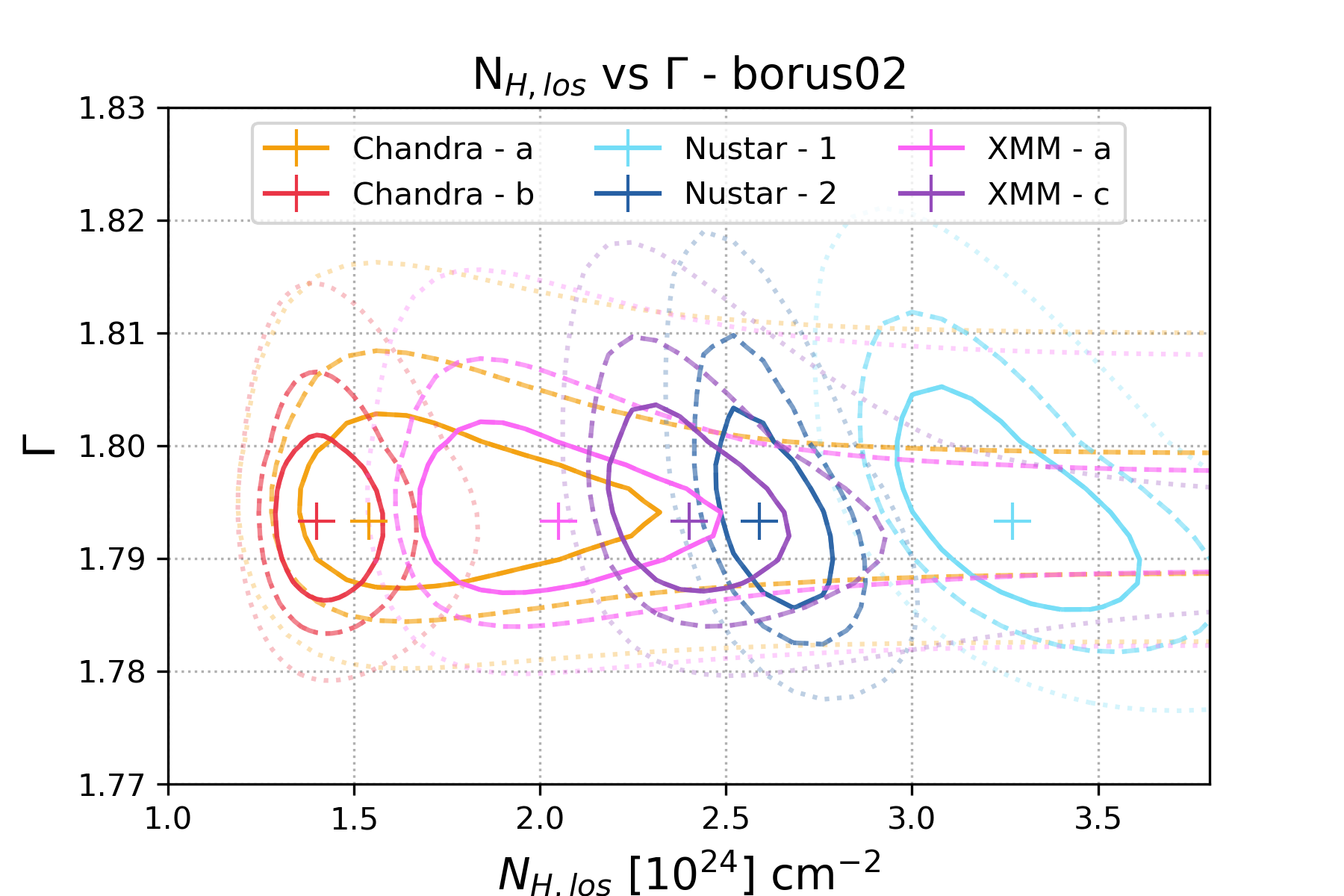}
    \caption{N$_{\rm H,los}$ vs cos$\theta_i$, (\textit{left}), and N$_{\rm H,los}$ vs $\Gamma$ (\textit{right}), multi-dimensional confidence contour plots obtained with \borus\ model for all observations. In both cases we see how all parameters are constrained withing $1\sigma$. }
    \label{fig:cont_nh_costheta_gamma}
\end{figure}

\begin{figure}[ht]
    \centering
    \includegraphics[ width=0.45\textwidth]{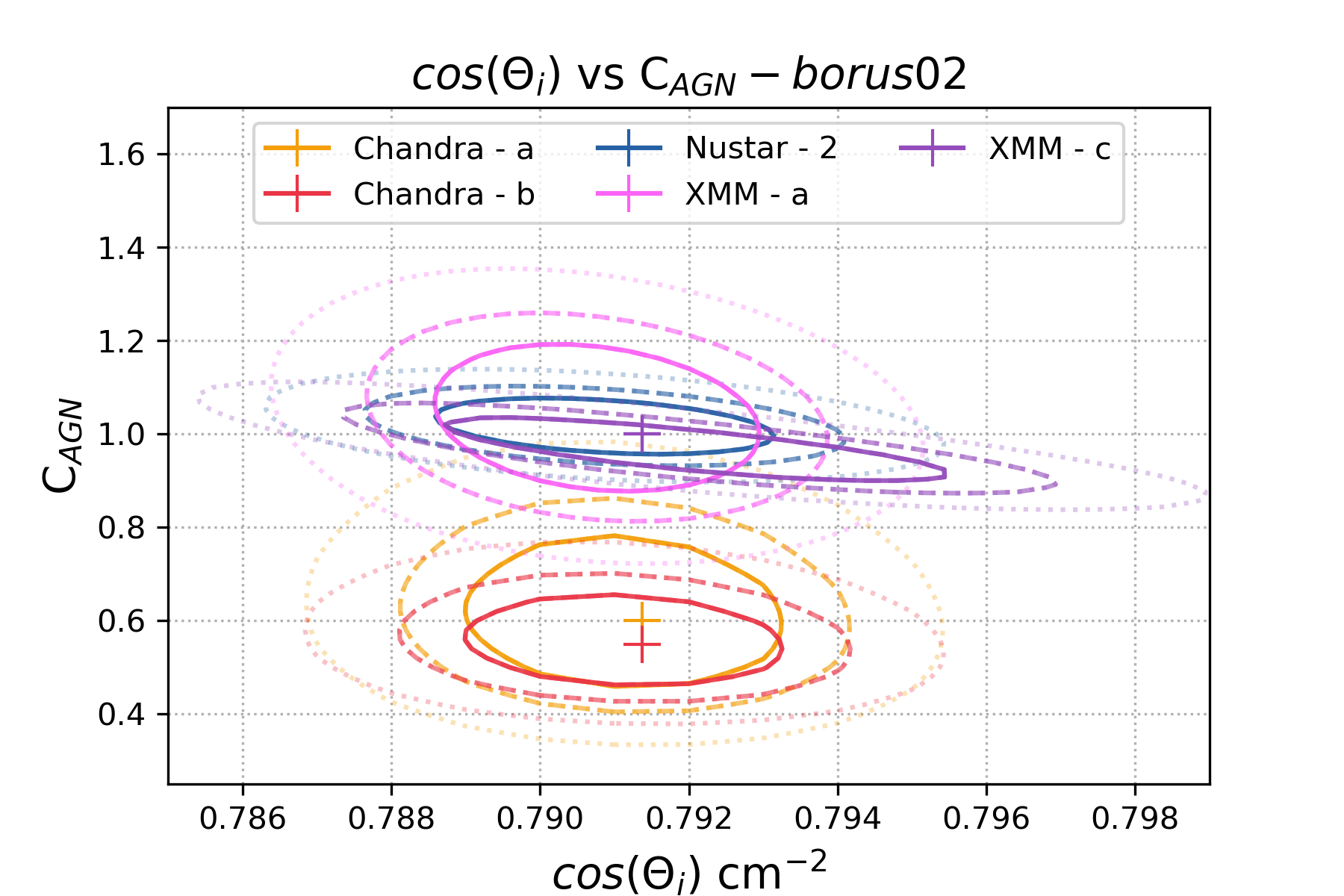}
    \includegraphics[ width=0.45\textwidth]{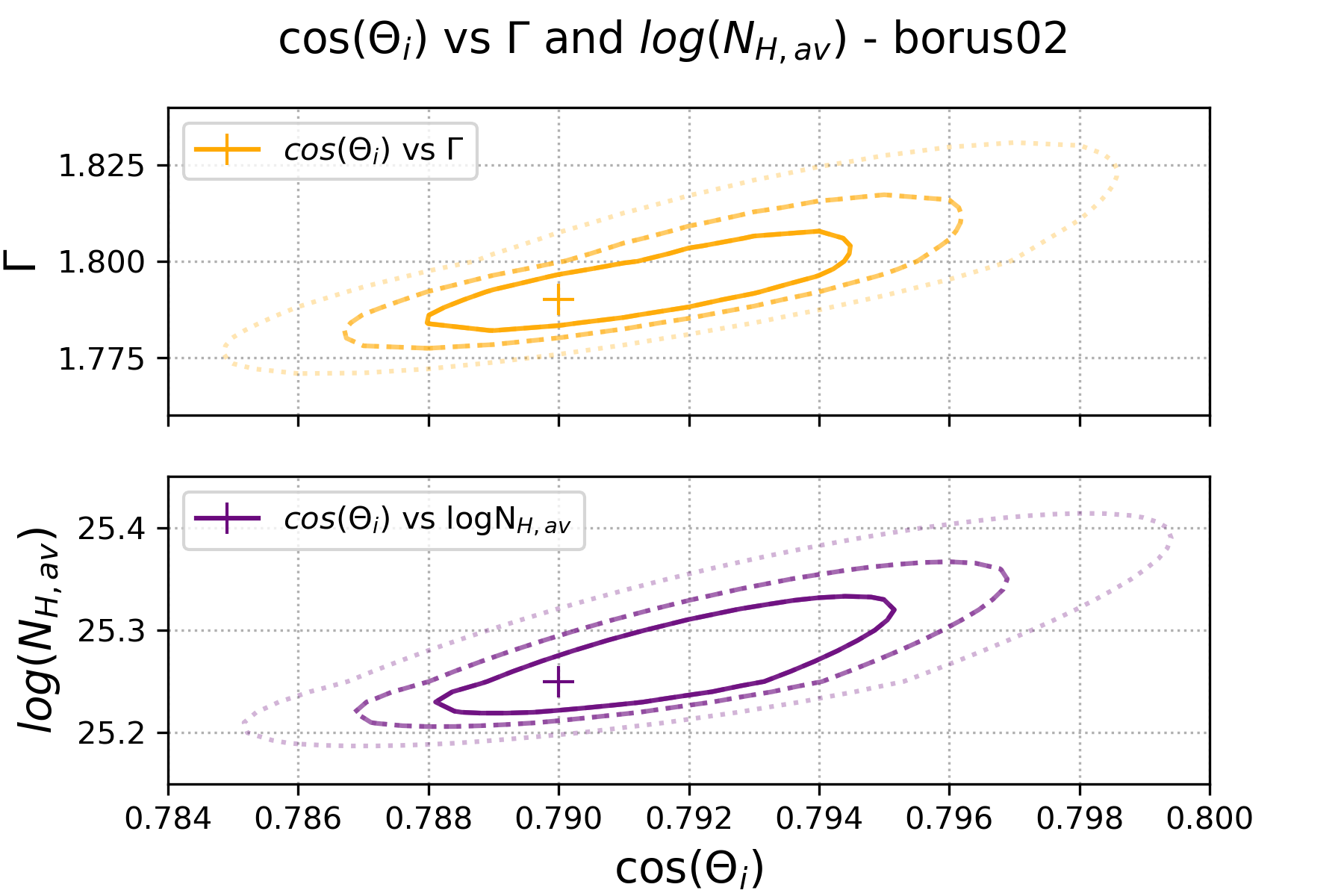}
    \caption{cos$\theta_i$ vs C$_{\rm AGN}$ (\textit{left}), and cos$\theta_i$ vs $\Gamma$ and log(N$_{\rm H,av}$)  multi-dimensional confidence contour plots obtained with \borus\ model for all observations. In this case, the degeneracy between the parameters is broken in all scenarios.  }
    \label{fig:cont_plot_costheta_c2_nhav_gamma}
\end{figure}

\end{document}